\PassOptionsToPackage{framemethod=tikz}{mdframed}
\documentclass[11pt,letterpaper,onecolumn]{arXiv-2605.19597v1/hunyuan_arxiv}

\graphicspath{{}{arXiv-2605.19597v1/}}

\usepackage[authoryear,sort&compress,round]{natbib}
\usepackage[most,breakable,skins]{tcolorbox}
\tcbuselibrary{skins,breakable}
\usepackage{microtype}
\usepackage{wrapfig}
\usepackage{listings}

\usepackage{booktabs}
\usepackage{tabularx}
\usepackage{array}
\usepackage{multirow}
\usepackage{tikz}
\usetikzlibrary{
    arrows.meta,
    positioning,
    fit,
    calc,
    backgrounds
}

\newtcblisting{toolcall}{
    listing only,
    listing options={
        basicstyle=\small\rmfamily,
        columns=fullflexible,
        keepspaces=true,
        breaklines=true,
        breakatwhitespace=false
    },
    colback=gray!5,
    colframe=gray!30,
    boxrule=0.4pt,
    arc=1pt,
    left=6pt,
    right=6pt,
    top=1pt,
    bottom=1pt,
    before skip=4pt,
    after skip=4pt
}

\let\cite\citep

\usepackage{amsmath,amsfonts,amssymb,amsthm}
\usepackage{booktabs,array,tabularx,multirow,ragged2e,xcolor}
\usepackage{tikz}
\usetikzlibrary{arrows.meta,positioning,shapes.geometric,fit,backgrounds,calc}
\usepackage[most]{tcolorbox}

\definecolor{exposeRed}{HTML}{B91C1C}
\definecolor{exploitAmber}{HTML}{B7791F}
\definecolor{misleadBlue}{HTML}{1D4ED8}
\definecolor{judgeBlue}{HTML}{1D4ED8}
\definecolor{softGray}{HTML}{F3F4F6}
\definecolor{borderGray}{HTML}{CBD5E1}
\definecolor{codeBg}{HTML}{F8FAFC}
\definecolor{codeFrame}{HTML}{D5DDE8}
\definecolor{codeTitleBg}{HTML}{EEF4FF}
\definecolor{codeTitleText}{HTML}{1E3A8A}
\definecolor{codeAccent}{HTML}{2563EB}
\definecolor{tableHeadBg}{HTML}{EAF1F8}
\definecolor{tableBandBg}{HTML}{F7FAFC}
\definecolor{tableGroupBg}{HTML}{F1F5F9}
\definecolor{tableRule}{HTML}{94A3B8}
\definecolor{tableText}{HTML}{1F2937}

\usepackage{enumitem}
\setlist[itemize]{leftmargin=*}
\setlist[enumerate]{leftmargin=*}

\theoremstyle{definition}

\AtBeginEnvironment{tabular}{\arrayrulecolor{tableRule}}
\AtBeginEnvironment{tabularx}{\arrayrulecolor{tableRule}}

\newtcblisting{boxedcode}[1][]{
  enhanced,
  listing only,
  listing engine=listings,
  breakable,
  width=0.88\textwidth,
  center,
  colback=codeBg,
  colframe=codeFrame,
  colbacktitle=codeTitleBg,
  coltitle=codeTitleText,
  fonttitle=\sffamily\bfseries\footnotesize,
  boxrule=0.45pt,
  arc=2pt,
  borderline west={2pt}{0pt}{codeAccent},
  left=9pt,
  right=9pt,
  top=7pt,
  bottom=7pt,
  before skip=8pt,
  after skip=10pt,
  listing options={
    basicstyle=\ttfamily\scriptsize\color{black!82},
    breaklines=true,
    columns=fullflexible,
    keepspaces=true,
    showstringspaces=false
  },
  #1
}

\AtBeginDocument{%
  \setlength{\parindent}{0pt}%
  \setlength{\parskip}{0.6\baselineskip plus 2pt}%
}

\makeatletter
\renewcommand{\abscontent}{}
\renewenvironment{abstract}
  {\begin{tcolorbox}[
      colback=gblue9!5,
      colframe=gblue9!5,
      boxrule=0pt,
      arc=6pt,
      left=8pt,right=8pt,
      top=6pt,bottom=6pt,
      breakable]
   \absfont}
  {\end{tcolorbox}\par\bigskip}
\makeatother

\usepackage{amsmath,amsfonts,bm}

\def\eqref#1{equation~\ref{#1}}

\def\1{\bm{1}}

\DeclareMathAlphabet{\mathsfit}{\encodingdefault}{\sfdefault}{m}{sl}
\SetMathAlphabet{\mathsfit}{bold}{\encodingdefault}{\sfdefault}{bx}{n}

\usepackage{fvextra}
\usepackage{hyperref}
\usepackage{longtable}
\usepackage{fontawesome5}
\usepackage{microtype}
\usepackage{booktabs}
\usepackage{pifont} 
\usepackage{multirow}
\usepackage{xspace}
\usepackage{color}
\usepackage{adjustbox}
\usepackage[edges]{forest}
\usepackage{amssymb}  
\usepackage{amsfonts}
\usepackage{tabularx}
\usepackage{tikz}
\usepackage{makecell}
\usepackage{caption}
\usepackage{subcaption}
\usepackage{amsmath}
\usepackage[utf8]{inputenc}
\usepackage[T1]{fontenc}
\usepackage{CJKutf8}

\DeclareUnicodeCharacter{2018}{`}%
\DeclareUnicodeCharacter{2019}{'}%
\DeclareUnicodeCharacter{201C}{``}%
\DeclareUnicodeCharacter{201D}{''}%
\DeclareUnicodeCharacter{2013}{--}%
\DeclareUnicodeCharacter{2014}{---}%
\DeclareUnicodeCharacter{2011}{-}%
\DeclareUnicodeCharacter{2026}{\ldots}%
\DeclareUnicodeCharacter{2022}{\textbullet}%
\DeclareUnicodeCharacter{25CF}{\textbullet}%
\DeclareUnicodeCharacter{2192}{\ensuremath{\rightarrow}}%
\DeclareUnicodeCharacter{00B7}{\textperiodcentered}%
\DeclareUnicodeCharacter{00BB}{\guillemotright}%
\DeclareUnicodeCharacter{00D7}{\ensuremath{\times}}%
\DeclareUnicodeCharacter{00B0}{\textdegree}%
\DeclareUnicodeCharacter{00B1}{\ensuremath{\pm}}%
\DeclareUnicodeCharacter{00F7}{\ensuremath{\div}}%
\DeclareUnicodeCharacter{00BA}{\textordmasculine}%
\DeclareUnicodeCharacter{00E9}{\'e}%
\DeclareUnicodeCharacter{200C}{}%

\usepackage{url}
\usepackage{nicefrac}
\usepackage{xcolor}
\usepackage[normalem]{ulem}
\usepackage[most]{tcolorbox}
\usepackage{colortbl}
\usepackage{tcolorbox}
\usepackage[tikz]{bclogo}
\usepackage{wrapfig}
\usepackage{float}
\usepackage{enumitem}
\usepackage{ifthen}
\ifdefined\directlua
  \IfFileExists{emoji.sty}{\usepackage{emoji}}{}%
\fi
\ifdefined\emoji\else
  \newcommand{\emojipenguin}{%
    \raisebox{-0.15\height}{\includegraphics[height=1.0em]{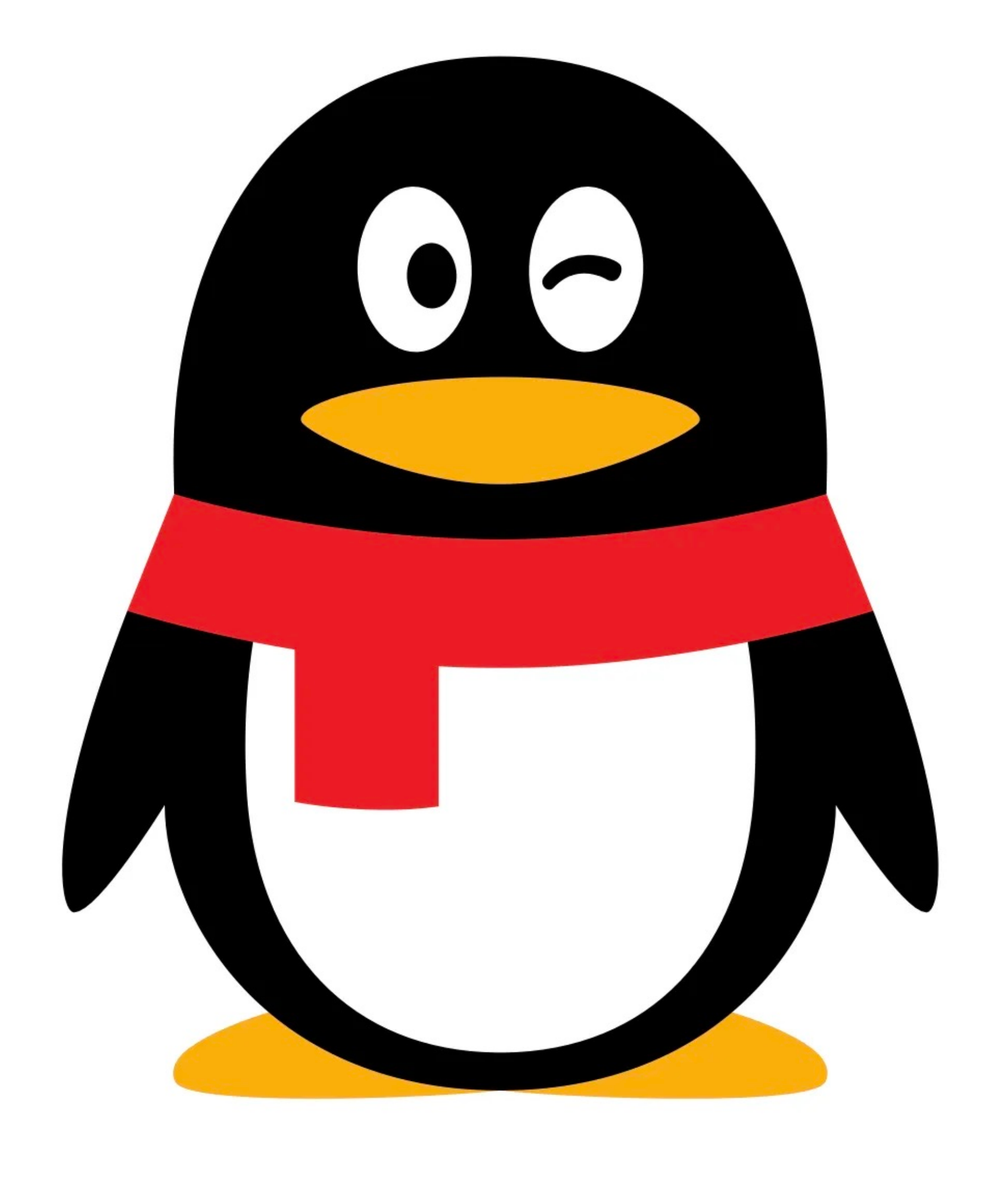}}%
  }%
  \providecommand{\emoji}[1]{\ifthenelse{\equal{#1}{penguin}}{\emojipenguin}{}}%
\fi

\usepackage{listings}
\usepackage{xparse}
\ExplSyntaxOn

\NewDocumentCommand{\gradientcell}{m}{
    \fp_set:Nn \l_tmpa_fp {#1}
    \fp_compare:nTF {\l_tmpa_fp > 20} {
        \cellcolor{green!60!yellow!\fp_eval:n{(\l_tmpa_fp-20)*2}!white}#1
    }{
        \fp_compare:nTF {\l_tmpa_fp > 10} {
            \cellcolor{yellow!60!red!\fp_eval:n{(\l_tmpa_fp-10)*3}!white}#1
        }{
            \cellcolor{red!\fp_eval:n{\l_tmpa_fp*5}!white}#1
        }
    }
}
\ExplSyntaxOff

\definecolor{deepred}{RGB}{230,101,101}

\definecolor{mygrey}{RGB}{105,105,105}

\usetikzlibrary{arrows.meta,positioning,fit,backgrounds,calc}

\renewcommand{\thefootnote}{\fnsymbol{footnote}}

\title{Multimodal Duplex Interaction Agent}

\makeatletter
\def\AB@authnote#1{}
\def\AB@affilnote#1{}
\renewcommand\AB@affilsepx{\protect\\ \protect\Affilfont}
\makeatother

\author{
Orantqing$^{*1}$ \quad Shengpeng Ji$^{*\dagger1}$ \quad Junlong Tong$^{*1,3}$ \quad Jialong Zuo$^{*1}$ \\ Dongjie Fu$^{1,2}$ \quad Di Cao$^1$ \quad Yangzhuo Li \quad Shangda Wu$^1$ \quad Franz \quad Evan \quad Theron Veyra \quad Changhao Pan$^2$ \quad Jingyu Lu$^2$ \quad Dongchao Yang$^4$ \quad Zhifei Xie$^5$ \quad Yang Tan \quad Xiaoyu Shen  \quad Xiaoda Yang$^2$ \quad Wenfu Wang$^1$ \quad Teddy Sun$^1$ \quad Steve Yves$^1$ \quad Zhou Zhao$^2$ 
}

\affil{$^1$Hunyuan Speech Team, Tencent; \quad $^2$Zhejiang University; \quad $^3$Shanghai Jiao Tong University}
\affil{$^4$The Chinese University of Hong Kong; \quad $^5$Nanyang Technological University}

\begin{document}

\footnotetext[0]{
\mbox{$^*$Equal contribution.\quad $^\dagger$Corresponding author.}\\
\mbox{\quad \quad This is an academic research project.\quad Email: \texttt{orantqing@gmail.com, shengpengji@zju.edu.cn}}
}

\maketitle

\vspace{-1em}

\begin{center}
    \begin{tabular}{@{}c@{}}
        \raisebox{-0.15\height}{\includegraphics[height=17pt]{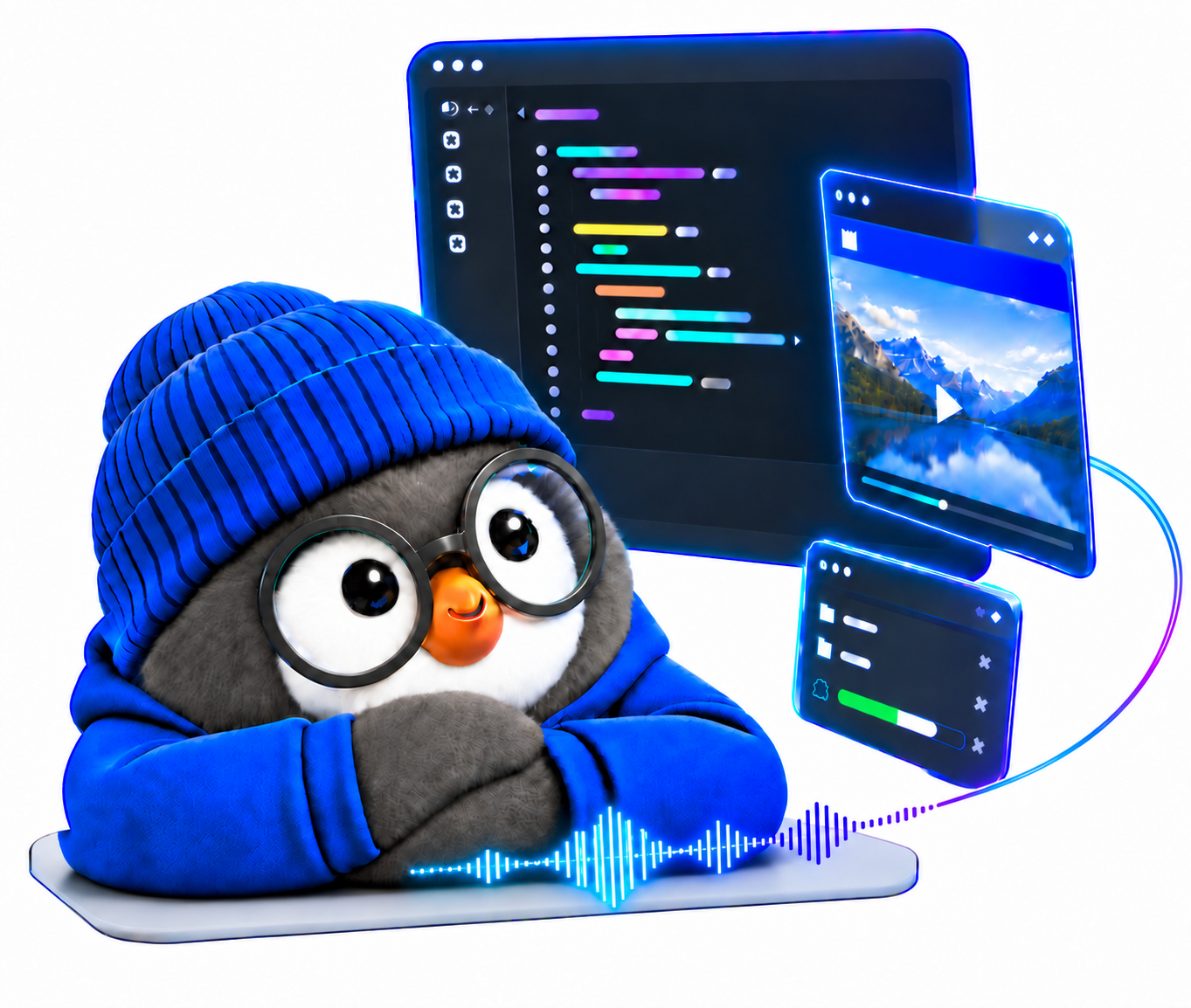}}\hspace{7pt}%
        \textbf{Research Project: }\href{https://Omni-Interaction-Gander.github.io/Omni-Interaction-Agent}{Gander Project Page}
        \\[6pt]
        \raisebox{-0.15\height}{\includegraphics[height=16pt]{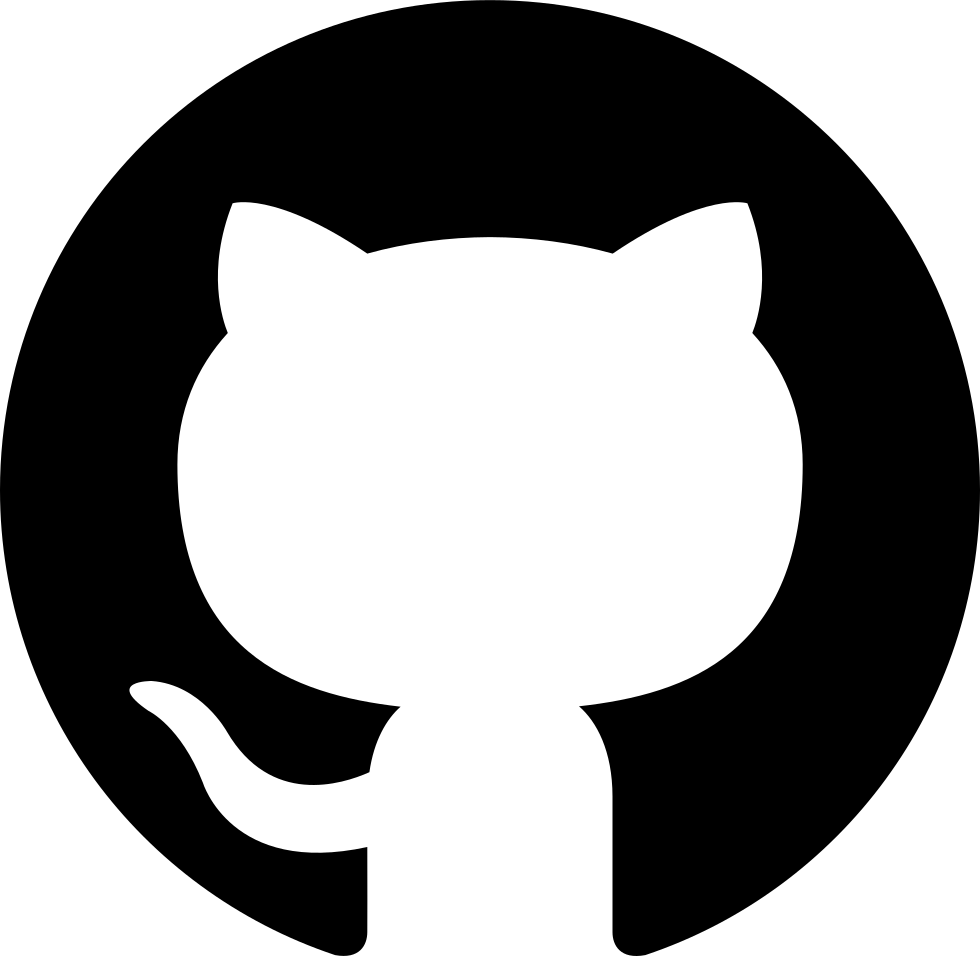}}\hspace{7pt}%
        \textbf{Code: }\href{https://github.com/Omni-Interaction-Gander/Omni-Interaction-Agent}{Gander Source Code}
    \end{tabular}
\end{center}



\begin{abstract}

In this work, we present \textbf{Gander}, \emph{a native multimodal duplex interaction model that builds on MiniCPM-o 4.5 and is further adapted for realtime interaction with an asynchronous agent loop}. In contrast to conventional turn based systems, \textbf{Gander} continuously processes streaming user inputs, enabling natural full-duplex interaction in both everyday conversations and complex workflow oriented agent scenarios. Users can interrupt an ongoing response, while the model can proactively provide intermediate feedback or ask follow up questions when appropriate. To natively support these capabilities, \textbf{Gander} adopts two key architectural designs: 1) \emph{a Cerebellum-Brain collaborative framework}, in which the Cerebellum is responsible for realtime interaction and conversational responsiveness, while the Brain handles complex reasoning and higher level agentic tasks. The two components interact continuously through \emph{tool calling and the agent orchestration runtime}. 2) The Cerebellum is built upon a streaming Thinker-Talker architecture, where \emph{user inputs and model outputs are flattened into an ordered token stream at the chunk level}, providing a unified representation for low latency and continuous interaction. We evaluate \textbf{Gander} across conversational ability, interactive capability, understanding, and tool assisted task execution. Internal human evaluations show that Gander maintains natural and expressive spoken dialogue, while benchmark results demonstrate effective turn taking capability and encouraging results on spoken question answering and related understanding tasks. Gander also supports a range of challenging interaction settings, including background noise interference, multi-party interactions, and backchannel communication. While our current evaluation focuses on representative conversational and tool assisted settings, broader long horizon agent tasks and more diverse deployment conditions remain promising directions for further study. We release \textbf{Gander} together with its models, code, and data to facilitate further research and development in the community.


\end{abstract}



\setcounter{footnote}{0}
\renewcommand{\thefootnote}{\arabic{footnote}}

\section{Introduction}

Large Language Models (LLMs) are rapidly evolving from language interfaces primarily used for question answering and conversation~\cite{yang2025qwen3} toward more general purpose agents capable of carrying out increasingly complex tasks~\cite{yao2022react, singh2025openai, anthropic2024claude}. By integrating tool use, environmental interaction, and multi step planning and execution, LLM based agents~\cite{glm5,kimik3} can go beyond generating textual responses to perceive, reason about, and act upon external environments. This emerging agentic paradigm has broadened the role of LLMs, enabling applications ranging from coding and tool assisted problem solving to workflow automation.

\begin{figure}[htbp]
\centering

\includegraphics[width=0.98\textwidth]{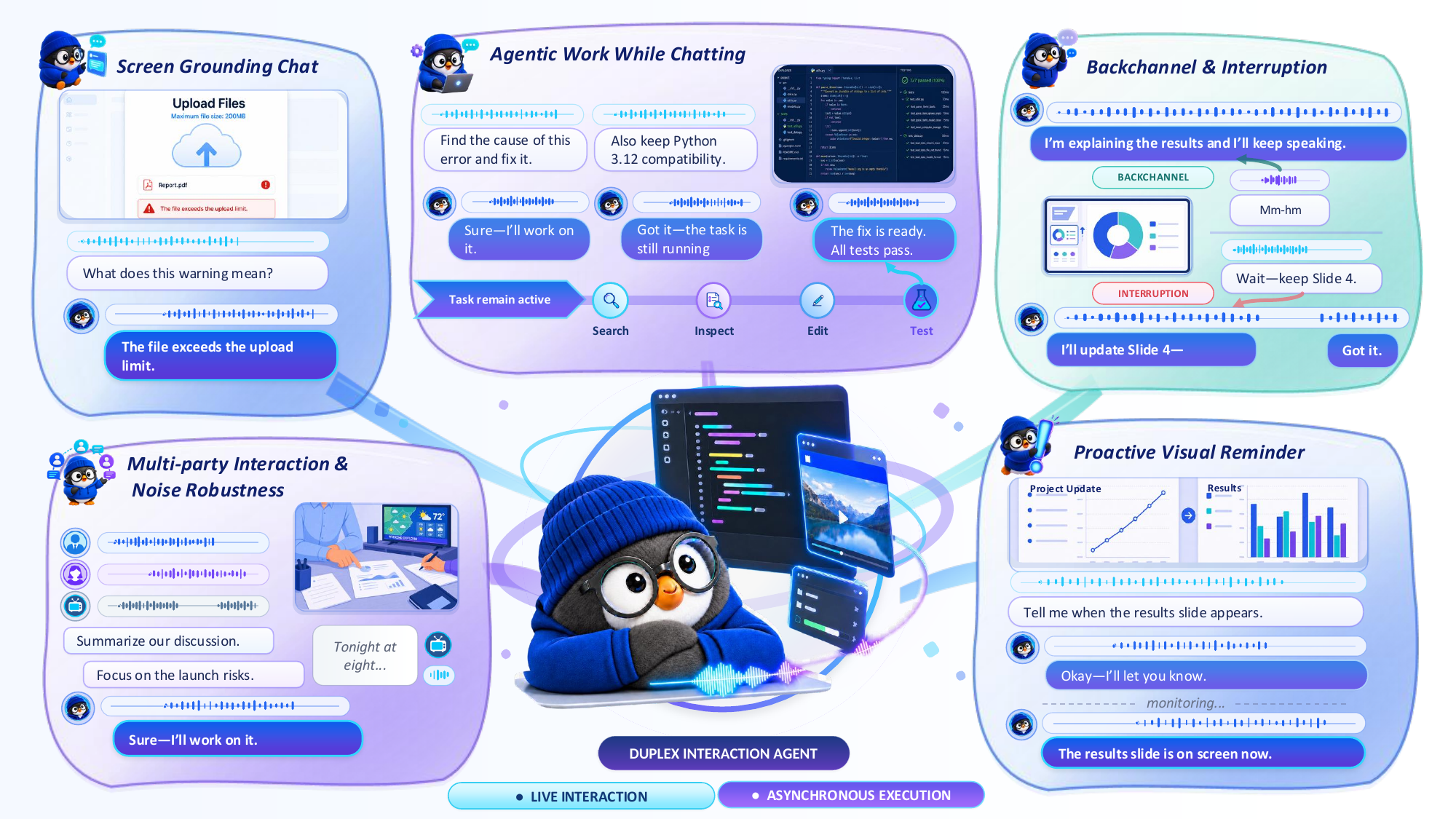}
\caption{
Gander combines realtime interactive communication with an asynchronous agent loop for multi step tasks in workflow oriented environments.
}
\label{fig1:example}
\vspace{-0.5em}
\end{figure}

Despite the rapid advancement of agentic capabilities, human-AI interaction is still commonly organized around \textbf{text based, turn-by-turn communication}, where users provide instructions and models respond in discrete interaction cycles. This paradigm differs from the fluid and collaborative nature of human-human communication, in which \textbf{participants continuously exchange information, listen and speak concurrently, observe contextual cues, and flexibly initiate, interrupt, or redirect an ongoing interaction}. A more natural and effective form of human-AI collaboration therefore motivates AI systems to move beyond conventional request response interaction and remain continuously engaged throughout an interaction. Realizing such a paradigm involves three complementary capabilities: 1) perception and understanding of information from speech, vision, and text; 2) continuous, low latency, and bidirectional \textbf{\textit{interaction}} that supports natural realtime communication; and 3) \textbf{\textit{agentic}} capabilities for contextual reasoning, planning, tool use, and task execution. Together, these capabilities provide a foundation for interactive agents that can perceive, communicate, reason, and act in dynamic environments while collaborating with humans on complex tasks.

Realizing the aforementioned form of human-AI collaboration raises two important questions. First, to what extent can natural realtime interaction be achieved by composing conventional perception and interaction modules, such as VAD~\cite{xu2026fireredasr2s} and ASR~\cite{whisper,funasr}, and to what extent should interactivity be modeled as an intrinsic capability of the model? Second, how can a system provide the low latency responsiveness required for realtime interaction while retaining the longer horizon reasoning needed for complex agentic tasks?

To address these questions, we present Gander, a duplex interaction model with an asynchronous agent loop designed to combine realtime multimodal interaction with agentic reasoning. 1) We argue that interactivity can benefit from being modeled within the interaction model rather than being handled entirely by an external orchestration pipeline. Real world human-AI interaction encompasses a range of dynamic conversational scenarios, including spontaneous user interruptions, agent initiated engagement, interaction under background noise, multi-party conversations, and backchannel behaviors (e.g., brief acknowledgments such as `uh-huh,'' `right,'' and ``I see''). Reliably coordinating these behaviors can be challenging for pipelines that rely on independently designed modules such as VAD~\cite{SileroVAD}. More importantly, \textbf{modeling interactivity as an intrinsic capability allows conversational control and the underlying model capability to be developed within a unified framework}. Inspired by this principle, Gander jointly processes streaming audio-visual inputs and the model's generated text stream by partitioning the streams into temporally aligned chunks and organizing them into a unified autoregressive sequence. Within each chunk, the model explicitly predicts whether to listen or speak, thereby learning to dynamically control its interaction state and coordinate perception and generation in realtime.2)  Realtime conversational interaction and complex agentic workflows impose different computational and reasoning requirements. Casual conversation favors immediate responses and continuous contextual adaptation, whereas complex workflows may involve longer horizon reasoning, iterative planning, tool use, and sustained execution. Handling both within a single monolithic model can create a tension between responsiveness and reasoning capacity. To address this challenge, following~\cite{thinkingmachines2026interactionmodels, huang2026duplexomni,Mind-paced-speaking}, \textbf{we adopt a Cerebellum-Brain style decoupled architecture, conceptually related to recent approaches that separate realtime interaction from asynchronous long horizon reasoning}. In Gander, the Cerebellum is responsible for realtime multimodal interaction, continuous perception, and responsive conversational control, while the Brain serves as a higher capacity reasoning module responsible for complex planning and agentic task execution. The Cerebellum continuously maintains the live interaction context and selectively invokes the Brain through tool calls, providing it with accumulated textual context, speech transcriptions, and visual frames. The Brain then performs deeper reasoning asynchronously and can return intermediate summaries, plans, or final results to the Cerebellum, which integrates these outputs into the ongoing interaction. Importantly, the Brain is designed as a plug-and-play component that does not require additional training within Gander, allowing different or stronger reasoning backends to be incorporated without retraining the core interaction model. This design provides a modular pathway for improving reasoning capability while preserving the realtime interaction model.

For reproducibility, we instantiate Gander by adapting MiniCPM-o 4.5~\cite{cui2026minicpm} as the initialization of the interaction model, and adopt a Thinker-Talker architecture related to recent streaming speech language systems~\cite{qwen3.5omni,qwen3omni}, together with a Codex based reasoning backend. Through training on conversational, interactive, multimodal understanding, and agentic data, as illustrated in Figure~\ref{fig1:example}, Gander demonstrates competitive performance across dialogue, multimodal understanding, realtime interaction, and tool assisted agentic tasks. It also shows useful context sensitive behaviors, such as resolving ambiguous references with visual context. To evaluate the complementary capabilities of duplex interaction agents, we follow evaluation practices used in recent realtime interactive systems~\cite{hurst2024gpt4o,openai_gpt_live_system_card_2026}, reporting results on established benchmarks including Full-Duplex-Bench, SpokenQA, and WorldSense, together with internal human evaluations of conversational naturalness and qualitative interaction examples on the project website. \underline{\textit{We release the data, code, and models to facilitate community research.}}

Our contributions are summarized as follows:

$(1)$ We introduce Gander, a duplex interaction model with an asynchronous agent loop that \textbf{supports modular extension of reasoning capability} and connects realtime interaction with longer horizon agentic reasoning.

$(2)$ Built on a Cerebellum-Brain collaborative framework with tool call feedback and chunk-based streaming, Gander demonstrates a broad set of interactive capabilities, including interruption, proactive interaction, robustness to background interference, multi-party interaction, backchannel communication, reasoning, and tool assisted task execution.

$(3)$ We release the Gander model weights, code, and data to facilitate open research and further development of duplex interaction agents.

\section{Related Work}


Interaction with speech language models has gradually evolved from conventional turn based human computer dialogue toward full duplex interaction~\cite{surveyduplex} and agentic task execution. Along this trajectory, existing research can be broadly organized into three closely related directions: audio interaction models, multimodal interaction models, and voice agent systems for practical task execution. Although these lines of research pursue different objectives, they share a common goal: moving beyond the conventional paradigm in which a model waits for a complete user utterance before responding, toward agents that can continuously perceive their environment, infer the current interaction state, and engage in communication in a natural and timely manner.

\subsection{Speech Language Models and Full Duplex Interaction} Early speech language models~\cite{ji2024wavchat,an2024funaudiollm,kimiaudio,stepaudio1} achieved substantial progress in speech understanding~\cite{qwenaudio,qwen2audio} and speech generation~\cite{cosyvoice3,controlspeech,ji2024mobilespeech,ji2025wavtokenizer,languagecodec}, but their interaction protocols were still largely governed by explicit dialogue turns. In typical systems, external modules such as Voice Activity Detection (VAD)~\cite{SileroVAD,xu2026fireredasr2s} are used to determine when the user begins and ends an utterance, after which the resulting speech segment is passed to the model for understanding and response generation. Consequently, the model itself has limited control over fundamental interaction decisions. Although this paradigm is effective for conventional question and answer interactions, it becomes less suitable for natural conversations involving hesitation, pauses, interruption, overlapping speech, and background interference. Recent work has therefore increasingly focused on enabling speech models to directly model interaction timing and conversational state.

BayLing-Duplex~\cite{BayLing-Duplex} incorporates decisions about when to listen, when to speak, and when to terminate the current response directly into a single autoregressive model. By introducing a small number of dedicated state tokens, the model is able to make interaction decisions during streaming speech generation without relying on an additional turn taking controller. This design treats interaction timing as part of the model's autoregressive prediction process rather than as an external system component. Qwen-Audio-3.0-Realtime~\cite{qwen_audio_realtime_2026} explores realtime speech interaction from a streaming perspective. Instead of treating each utterance as a complete acoustic segment, the continuous audio stream is processed incrementally \textbf{in chunks}, allowing the model to continuously acquire contextual information and generate responses with reduced interaction latency. Audio Interaction Model~\cite{AudioInteractionModel} extends the scope of realtime interaction from conversational speech to general online audio interaction. The Model continuously listens to environmental sounds and user instructions and determines whether a response is necessary. The emphasis on proactive interaction is particularly important because the model is no longer required to respond only after an explicitly defined user turn. Instead, it can initiate a response when the semantics of the ongoing audio stream indicate that intervention is appropriate. Seeduplex~\cite{seeduplex2026} further investigates the challenges of continuous listening in realistic acoustic environments. Its focus goes beyond enabling simultaneous listening and speaking to include attentive listening and robust interference suppression. In particular, the model is required to distinguish relevant user speech from background sounds and speech produced by other speakers while maintaining the ability to respond appropriately to user interruptions. These capabilities highlight an important aspect of full duplex interaction: natural interaction depends not only on simultaneous input and output, \textbf{but also on the model's ability to maintain an appropriate interaction state under complex acoustic conditions.} GPT-Live~\cite{openai_gpt_live_system_card_2026} represents another important line of development toward natural realtime speech interaction. GPT-Live adopts a full-duplex architecture similar to Moshi~\cite{defossez2024moshi}, allowing it to continuously process incoming audio and generate speech output in parallel. This allows the system to make interaction decisions during an ongoing conversation, including whether to continue listening, respond, pause, or interrupt. 

These studies have substantially advanced speech interaction beyond the conventional turn based paradigm. The central question has shifted from whether a model can understand and generate speech to whether it can continuously listen, speak, and regulate its own participation in a conversation. Nevertheless, most of these efforts remain centered on audio based interaction. Their understanding of the surrounding environment is therefore primarily derived from acoustic information, and their interaction capabilities are still largely evaluated within relatively constrained conversational settings.

\subsection{Multimodal Models and Continuous Interaction} The integration of visual, audio, and video information into a unified foundation model has opened a new research direction toward multimodal interaction. Compared with speech only systems, multimodal models have access to both auditory and visual context, enabling the model to reason about not only what the user says, but also the surrounding scene and its temporal evolution. This additional context provides new opportunities for interaction, particularly in situations where the meaning of speech depends on visual information. Representative models such as Qwen Omni~\cite{qwen3.5omni,qwen3omni} series integrate multiple input modalities within a unified foundation model and support streaming speech generation. These systems provide a general foundation for realtime multimodal interaction by enabling the model to jointly interpret linguistic, acoustic, visual, and temporal information within a shared context.

MiniCPM-o 4.5~\cite{cui2026minicpm} further explores continuous multimodal interaction through Omni Flow. Rather than treating multimodal perception and response generation as independent stages, Omni Flow places multimodal inputs and outputs on a unified temporal axis, allowing the model to continuously perceive visual and audio signals while generating responses. This design is particularly relevant to realtime interaction because the model can maintain a continuous temporal representation of the interaction rather than repeatedly resetting its context at individual turns. MiniCPM-o 4.5 demonstrates proactive behaviors based on continuously observed environmental information. JoyAI-VL-Interaction~\cite{yao2026joyai} focuses on interaction over continuous visual streams. Instead of treating vision as a static source of contextual information, it considers continuous visual perception as part of the interaction process itself. The model observes changes in the environment over time and determines whether these changes warrant a response or further interaction. SeedRealtime~\cite{seedrealtime2026} further advances this direction by introducing native audio visual full duplex interaction. It jointly integrates audio, video, and text within a unified architecture and enables continuous interaction over multimodal streams. Beyond simply combining multiple input modalities, SeedRealtime demonstrates that audio visual integration can directly improve interaction quality. For example, visual context can help resolve phonetic ambiguity, interpret temporal references, and connect what is being seen, heard, and said within the same interaction process. The model also demonstrates proactive interaction by responding to changes in the observed environment without requiring an explicit user request.


\subsection{From Conversation to Voice Agents}

Although the above approaches have improved the naturalness of conversational interaction, practical deployment introduces a further challenge. Users increasingly expect AI systems not only to converse with them, but also to perform meaningful tasks, such as writing code~\cite{glm5}, accessing files, invoking external tools~\cite{voxmind}, and executing multi step workflows. In these settings, conversational interaction and task execution are closely coupled, yet they are still commonly implemented as partially independent capabilities.

Qwen Audio Agent~\cite{qwen_audio_agent_2026} represents a system oriented toward this problem. It provides a realtime voice runtime that enables an agent to maintain continuous spoken interaction while delegating more complex tasks to a backend workflow. Such tasks can include code generation, computer file operations, and other work related activities. This design moves voice interaction beyond a conversational interface and toward a persistent interface to an agent workflow. GPT-Live with CodeX~\footnote{\url{https://help.openai.com/en/articles/20001274-chatgpt-voice?utm_source=chatgpt.com}} and Claude Voice Mode~\footnote{\url{https://support.claude.com/en/articles/11101966-use-voice-mode?utm_source=chatgpt.com}} demonstrate product oriented approaches in which natural voice interaction is integrated with general purpose AI assistant capabilities. These systems make increasingly complex AI functionality accessible through continuous spoken interaction, thereby reducing the distinction between a conversational interface and a general AI assistant.

Despite these advances, current voice agent systems still face substantial challenges in realistic interactive workflows. Throughout task execution, users may interrupt the agent and provide new constraints, ask follow up questions, or correct previously stated information. The agent may also need to proactively communicate intermediate results, solicit clarification, and request additional details to ensure that the task proceeds correctly. A brief backchannel does not necessarily indicate that the current task should terminate. In multi party environments, the system may need to distinguish relevant speech from other speakers while simultaneously handling background noise and changes in the visual environment. These scenarios require the model to reason jointly about interaction state, environmental state, and task state. Treating speech only as an input and output interface is therefore insufficient for truly natural agent interaction.

Motivated by these developments, we propose \textbf{a duplex interaction model called Gander} which studies a form of interaction in which an end-to-end interaction model supports natural realtime conversation while coordinating asynchronous agentic task execution. The key question is whether conversational interaction, multimodal perception, and task execution can be integrated into a single continuous process rather than being handled as separate interaction stages. Gander brings these capabilities together through a duplex interaction model and an agent loop. Rather than focusing solely on conversational interaction, Gander is designed to move naturally between open ended conversation and practical task execution. Rather than treating speech as merely an interface to an external agent, Gander incorporates interaction state into the end to end modeling process, allowing conversational behavior and task execution to be coordinated within the same ongoing interaction.

\section{Gander: Duplex Interaction and Agent Loop} \label{sec:e_bench}


\subsection{Overview} \label{subsec:e_bench_overview}

Gander couples a duplex interaction model with an asynchronous agent loop and is architecturally composed of \textbf{\textit{the front cerebellum, the agent orchestration runtime, and the back brain}}. The front cerebellum is a realtime, full duplex multimodal model based on a thinker-talker architecture, responsible for continuous multimodal perception and interactive communication. The back brain is a general purpose task execution agent that requires no additional task specific training within Gander, instantiated by general purpose agents such as claude code and codex. The agent orchestration runtime serves as the coordination layer between the front cerebellum and back brain, jointly managing realtime multimodal inference and the orchestration of asynchronous background tasks. Through this division of responsibilities, the three components collectively support both continuous realtime interaction and complex long horizon task execution. We detail their respective architectures and interactions below.

The remainder of this section is organized as follows. Section~\ref{subsec:env-construction} describes the interaction between the brain and cerebellum, including \textbf{\textit{front cerebellum}} activation and input transmission, the \textbf{\textit{intermediate agent orchestration runtime}}, and the propagation of responses from the \textbf{\textit{back brain}}. Section~\ref{subsec:task-construction} details the front cerebellum architecture, including its Thinker-Talker design and streaming chunk flattening mechanism for multimodal interaction. 

\subsection{Cerebellum-Brain Collaborative Framework} \label{subsec:env-construction}

\begin{figure}[t]
\centering
    \includegraphics[width=0.98\textwidth]{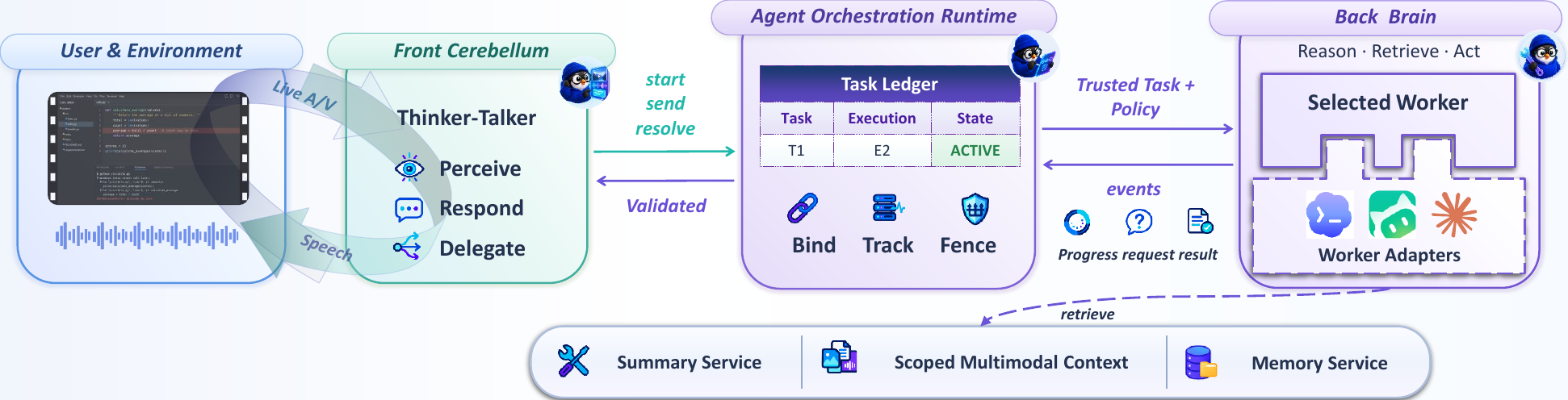}
\caption{
Gander consists of three components: the front cerebellum, the agent orchestration runtime, and the back brain. The front cerebellum handles realtime user interaction, while the back brain performs complex reasoning and long horizon workflow execution. The agent orchestration runtime serves as the coordination layer for realtime multimodal inference and asynchronous background task orchestration.
}
\label{fig1:shengpengpic2}
\end{figure}

Gander employs a decoupled, two tier agent architecture consisting of a front cerebellum and a back brain. As shown in Figure~\ref{fig1:shengpengpic2}, the front cerebellum is instantiated as a realtime, full-duplex multimodal model that serves as the primary interface \textbf{for continuous interaction, processing streaming speech, camera, and screen inputs and autoregressively determining interaction actions, including listening, speaking, brief feedback, interruption, and tool invocation}. The back brain is instantiated as a general purpose task execution agent for information retrieval, code and file manipulation, document processing, and other long horizon tasks. The intermediate agent orchestration runtime provides the execution substrate that coordinates the two layers, supporting unified realtime multimodal inference and asynchronous background task orchestration, and constitutes an integral component of the overall agent harness. We next characterize \underline{\textit{the information flow across these components}} from three complementary perspectives.

\subsubsection{Front Cerebellum Activation and Summarization}

The front cerebellum is itself a realtime duplex interaction model, broadly analogous to systems such as GPT-Realtime~\cite{openai_gpt_live_system_card_2026} and Seed-Realtime~\cite{seedrealtime2026}, and can independently handle routine conversational exchanges, information seeking interactions, and simple search based tasks. Within the dual brain architecture, \textbf{\textit{it additionally assumes the role of task routing and back brain invocation}}. Based on the current interaction context, the front cerebellum dynamically determines whether a request can be resolved locally or requires delegation to the back brain. Simple conversational and short horizon tasks are handled directly by the front cerebellum, whereas complex workflows involving multi step reasoning, external tool use, or long horizon execution are delegated to the back brain.

In Gander, this delegation mechanism is formalized through structured tool calls that expose the front cerebellum's task-level decisions to the agent orchestration runtime. Specifically, the front cerebellum emits tool invocations enclosed by dedicated special tokens:

\begin{toolcall}
<tool_call> {"name": "tool_state", "arguments": {"key": "value"}} </tool_call>
\end{toolcall}

\begin{figure}[t]
\centering
    \includegraphics[width=0.98\textwidth]{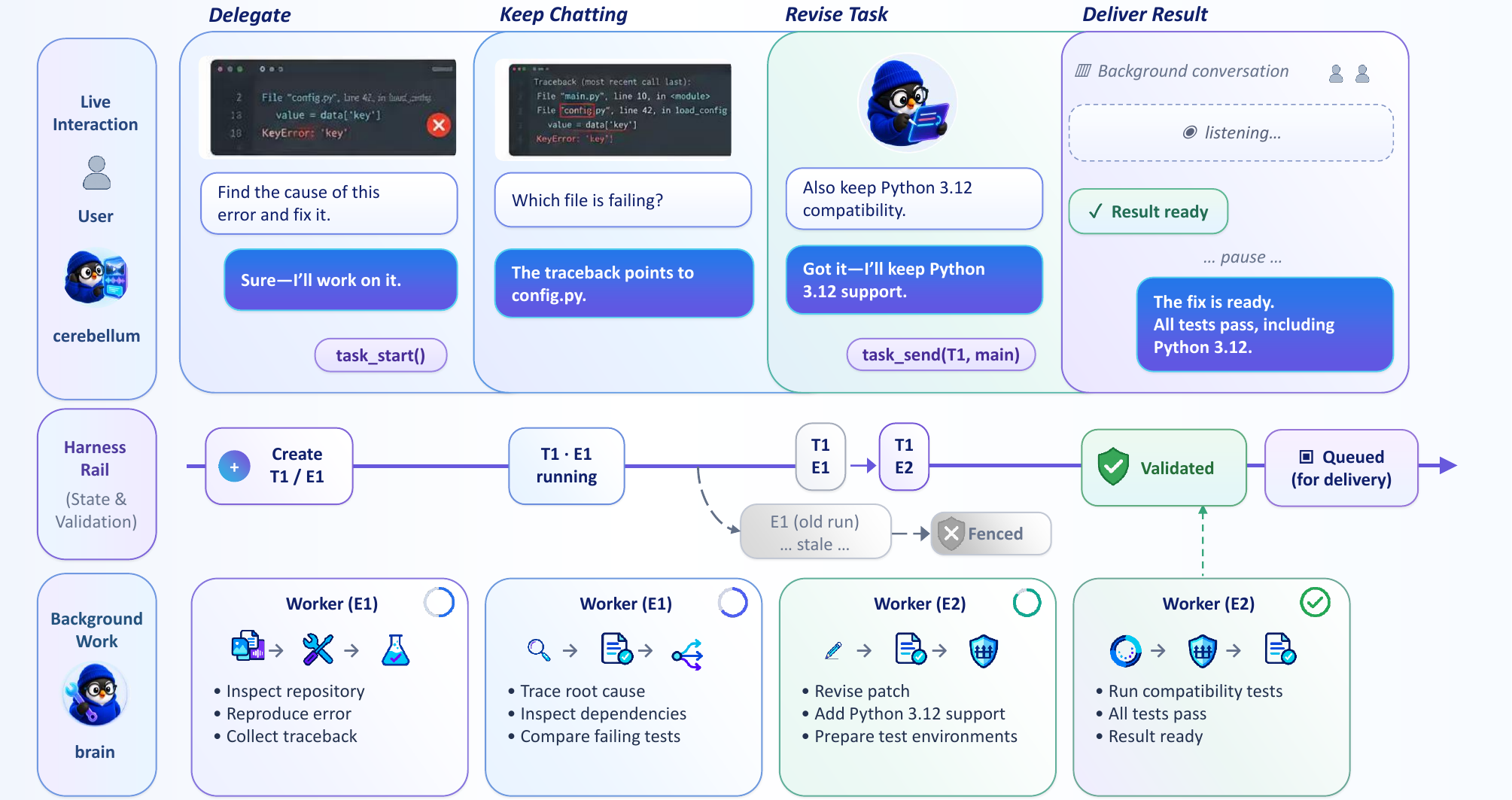}
\caption{
An example workflow illustrating cerebellum brain interaction, where users can engage in realtime conversation or modify previously assigned tasks while the back brain is executing ongoing workflows.
}
\label{fig1:shengpengpic3}
\end{figure}

The interface defines three primary operations: \textbf{\textit{\underline{task$\_$start}, \underline{task$\_$send}, and \underline{task$\_$resolve}}}, corresponding to task creation, incremental task interaction, and deterministic \textbf{\textit{\underline{tool$\_$state}}} control respectively in the above tool calls. The three tool states are defined as follows.

\begin{itemize}
    \item \textbf{\textit{task$\_$start.}} It initializes a new background task and associates it with the current user turn. The provided task description or name serves as the task identifier for subsequent lifecycle management, tracking, and interaction.

    \item \textbf{\textit{task$\_$send.}} It routes subsequent user inputs to an existing task, enabling incremental task specification and dynamic task refinement. It is invoked when the user provides additional information, revises a prior request, or otherwise modifies an ongoing task. The operation supports two routing modes: 1) \texttt{main}, which steers the primary task by incorporating the new input into its active execution context; 2) \texttt{fork}, which instantiates a read only auxiliary inquiry without altering the state or execution trajectory of the primary task.

    \item \textbf{\textit{task$\_$resolve.}} It provides deterministic control over the task lifecycle through four resolution actions: \texttt{cancel}, \texttt{allow$\_$once}, \texttt{allow$\_$session}, and \texttt{deny}. \texttt{cancel} terminates the active primary task; \texttt{allow$\_$once} authorizes the current operation on a one time basis; \texttt{allow$\_$session} extends the authorization to the duration of the current task session; and \texttt{deny} rejects the requested operation and blocks its execution. This interface enables the front cerebellum to express task level control decisions in a structured and explicit form, while the agent orchestration runtime deterministically enforces the associated state transitions and execution semantics.
\end{itemize}

In Figure~\ref{fig1:shengpengpic3}, the front cerebellum and back brain workflow is illustrated through a representative use case. Here, $T$ denotes a task, while $E$ denotes the back brain execution. The front cerebellum initiates a back brain tool call via \textbf{\textit{task$\_$start.}} During back brain execution, users can engage in conversation with the front cerebellum or modify previously issued tasks through \textbf{\textit{task$\_$send.}}

For the various tool calling scenarios triggered by the front cerebellum described above, the front cerebellum forwards the transcribed text of the audio query together with the relevant final frames of the input video, to the intermediate agent orchestration runtime as the input to the back brain. This design presents several open design considerations, including whether the front cerebellum should perform additional reasoning based on the transcribed query before passing it to the subsequent stage, and whether a trainable back brain could directly receive the user's complete multimodal input stream in real time. In Gander, we adopt the most straightforward approach, directly using the transcribed query and the relevant final video frames as the input to the back brain.

Within the two-brain architecture, the front cerebellum is also responsible for \textbf{summarizing the results} returned by the back brain and generating the final response to the user. This involves different scenarios, such as whether the summary should be based on the back brain's intermediate execution states or its final result. We treat this as a data scaling design choice, determined by the training data and target interaction scenarios. To preserve the interactive nature of multimodal interaction, the synthesized response is optimized for natural conversational spoken language delivery, rather than being formulated solely as a textual task completion response.


\subsubsection{Agent Orchestration Runtime}
The agent orchestration runtime primarily serves as the coordination layer between the front cerebellum and the back brain. By providing a unified interface for realtime multimodal data transport and backend task orchestration, the runtime effectively turns the duplex interaction agent into a unified harness for interactive task execution. For realtime interaction, \textbf{\textit{it manages audio and visual stream ingestion, incremental inference scheduling, serialization of model invocations, static prefix caching, and source aware rate limiting with persistent state across different visual inputs}}. The runtime also binds the \textbf{\textit{task$\_$start}}, \textbf{\textit{task$\_$send}}, and \textbf{\textit{task$\_$resolve}} operations issued by the front cerebellum to the final user turn confirmed at the transport layer, ensuring that task objectives are consistently grounded in authenticated user input rather than tool parameters independently generated by the front cerebellum.

The agent orchestration runtime supports two control modes: \texttt{lean} and \texttt{coordinator}.

$(1)$ In the \texttt{lean} mode, the runtime directly executes the task actions classified by the front cerebellum and uses the original user turn as the instruction to the back brain. This results in a shorter control path, lower additional latency, greater determinism, and fewer potential model failure points, making it suitable for scenarios in which the front cerebellum provides reliable task routing. Its limitation is that task configuration is largely determined at deployment time, making it difficult to dynamically adapt the reasoning intensity, supervision policy, or notification granularity to the semantics of individual tasks. In addition, routine intermediate events are more likely to be propagated directly through the delivery pipeline.

$(2)$ The \texttt{coordinator} mode introduces an independent control plane model between the front cerebellum and the gateway. Conditioned on trusted user requests and bounded system state, the \texttt{coordinator} generates execution directives that specify the reasoning intensity, questioning policy, permission policy, and result delivery strategy. This design decouples complex task planning and supervision decisions from the realtime front cerebellum, at the cost of additional model invocations, latency, computational overhead, and nondeterminism. The \texttt{coordinator} only produces a declarative control plan and does not execute user tasks or access files, shell, or network resources. Its output is subsequently subjected to structured validation by the gateway (the core control component of the runtime) against the current task state, provider capabilities, permission boundaries, and deployment configuration. In the current implementation, the \texttt{coordinator} is primarily involved in \textbf{\textit{task$\_$start}}, whereas \textbf{\textit{task$\_$send}} and \textbf{\textit{task$\_$resolve}} are handled directly by the gateway.


The Gateway serves as the persistent orchestration core of the Runtime. As shown in Table~\ref{tab:runtime_entities}, it organizes backend execution around five persistent entities, namely \textit{Project}, \textit{Task}, \textit{Run}, \textit{WorkerEvent}, and \textit{Delivery}. It manages task state transitions, worker scheduling, workspace isolation, concurrency control, permission handling, and result delivery. Different back brains are integrated through a unified \textit{Worker Provider} interface, with each \textit{Provider} declaring its supported capabilities, including native steering, read only side queries, user interaction, cross process recovery, supported input modalities, context acquisition mechanisms, and parallel execution. The current default back brain is implemented using the Codex app server, with each task lineage associated with a persistent Codex thread that can be incrementally updated throughout execution. User follow up instructions can be injected directly into the active thread, whereas read only side queries are executed in isolated forks to preserve the state of the primary thread and workspace.

\begin{table}[t]
\centering
\caption{Core entities managed by the gateway in the agent orchestration runtime.}
\label{tab:runtime_entities}
\begin{tabular}{ll}
\toprule
\textbf{Entity} & \textbf{Description} \\
\midrule
\textit{Project} & A persistent container for a long-lived task or workflow. \\
\textit{Task} & A logical unit of work representing a user's intended objective. \\
\textit{Run} & A concrete execution instance of a task. \\
\textit{WorkerEvent} & An event emitted by a worker during task execution. \\
\textit{Delivery} & The mechanism and record through which task results are delivered to the user. \\
\bottomrule
\end{tabular}
\end{table}

\subsubsection{Training Free Back Brain}

The back brain is a general purpose task execution agent integrated without additional training, instantiated by systems such as Codex or Claude Code. It is responsible for long horizon reasoning and sustained task execution, including information retrieval, code and file manipulation, document processing, and other tool mediated workflows. In addition to its native file, command line, and retrieval tools, the back brain can invoke three runtime specific interfaces: 1) \texttt{context$\_$fetch}, which retrieves realtime task context, events, and artifacts from the current task lineage based on references, roles, event types, time ranges, and character budgets; 2) \texttt{memory$\_$search}, which retrieves persistent multimodal memories across sessions; 3) \texttt{share}, which returns verified important findings, intermediate progress, or corrective information to the realtime front cerebellum.

\subsection{Streaming Thinker-Talker Architecture} \label{subsec:task-construction}

\begin{figure}[t]
\centering
    \includegraphics[width=0.98\textwidth]{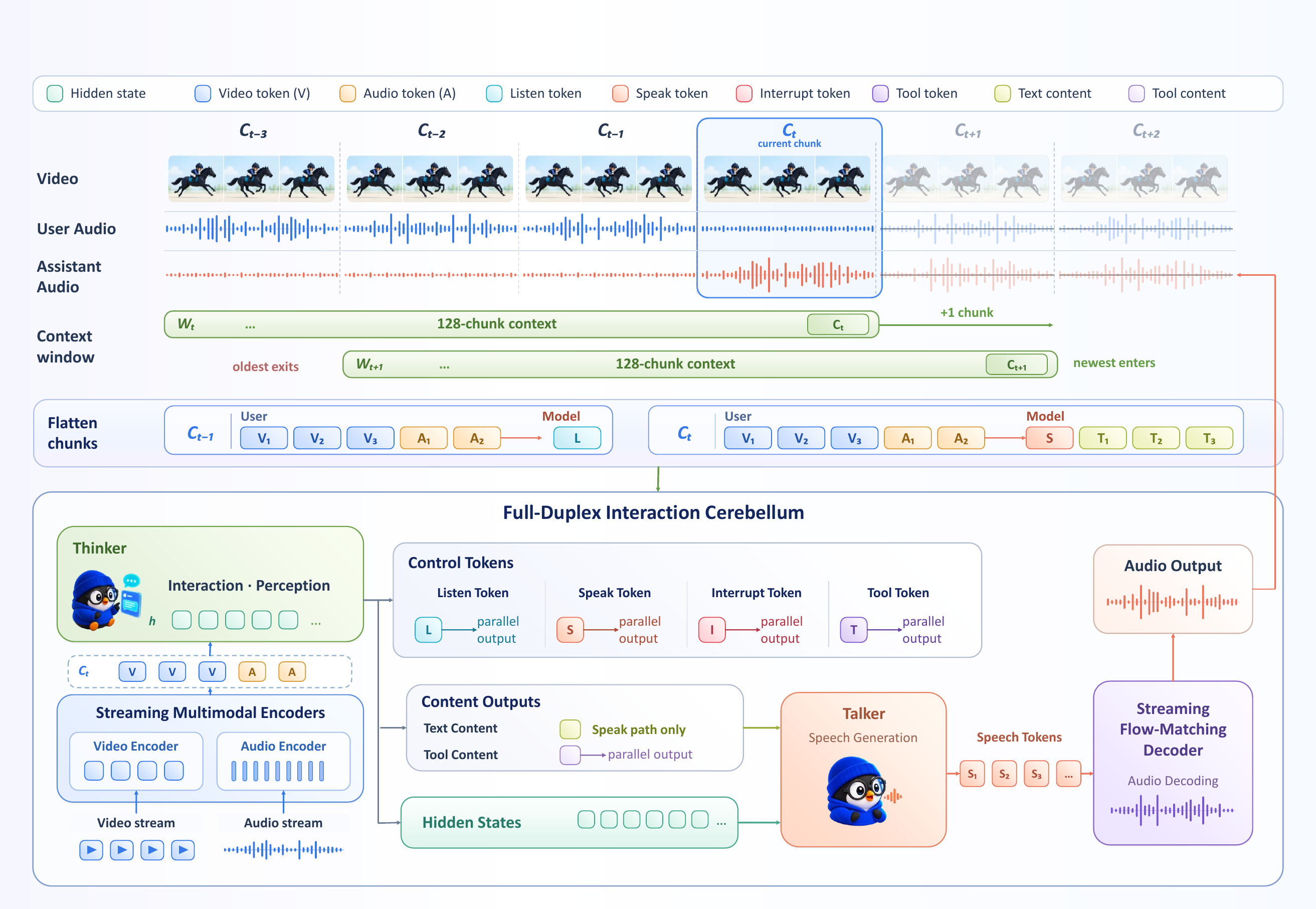}
\caption{
Detailed architecture of the front cerebellum. The front cerebellum adopts a classic thinker-talker design that ingests streaming audio-visual input and produces both text and speech. The LLM flattens inputs and outputs into a unified chunk stream, predicting a control token within each chunk to decide its interaction behavior, and maintains a lightweight context via a sliding window.
}
\label{fig1:shengpengpic4}
\end{figure}

In this section, we detail Gander's front cerebellum and its multimodal interaction capabilities. As illustrated in Figure~\ref{fig1:shengpengpic4}, the front cerebellum builds on a thinker-talker architecture~\cite{cui2026minicpm} that receives streaming audio and video input and 
generates both textual and audio output. Building on this design, we introduce a streaming chunk flattening mechanism for end-to-end interaction, which flattens perceptual inputs and generated outputs into a unified chunk stream and enables the model to dynamically determine whether to listen or speak at each chunk.

\subsubsection{Multimodal Perception} \label{jsp:1}

The front cerebellum senses its environment through two modality encoders that run concurrently, one over the visual stream and one over the acoustic stream, and projects their outputs into the token space of the LLM backbone so that all perceptual evidence is registered on the shared interaction timeline. Neither encoder waits for a complete clip or utterance before yielding representations. Both operate incrementally over fixed duration windows, so that visual and auditory observations are continuously refreshed and exposed to the backbone at every step of the streaming exchange rather than only at turn boundaries. We describe the two perception pathways below.

\vspace{-0.5cm}

\paragraph{Visual perception.} To ingest frames of arbitrary aspect ratio and resolution while keeping the token budget affordable under realtime constraints, the visual pathway follows an any resolution partitioning scheme~\cite{llavauhd}: each frame is first decomposed into a set of slices, every slice is independently encoded by a SigLIP vision transformer~\cite{siglip}, and the resulting patch features are then condensed by a query based resampler~\cite{minicpmv} into a small, fixed number of visual tokens per slice. This yields roughly a $16\times$ reduction relative to the raw patch grid, markedly more aggressive than the $4\times$ compression adopted in many prior multimodal models, and substantially lowers the per-frame cost of continuous perception. Since full-duplex interaction imposes a stringent latency profile in which frames arrive densely and must be consumed within tight time windows, the cerebellum operates at a compact resolution of up to $448\times448$, which keeps the per-frame token count small enough to sustain continuous perception in real time.

\vspace{-0.5cm}

\paragraph{Acoustic perception.} The acoustic stream, which conveys both the ambient scene and user speech, is processed by a streaming and chunk wise speech encoder~\cite{whisper} that emits a dense sequence of frame level features on the order of $50$ frames per second. Passing this rate directly to the backbone would inflate the sequence length and undermine the efficiency that realtime operation demands; we therefore interpose a lightweight MLP projector that applies a $5\times$ temporal downsampling, lowering the effective audio token rate to roughly $10$ tokens per second before the features enter the backbone. This compression retains the phonetic and prosodic cues required for accurate spoken-language understanding while keeping the audio stream commensurate with the backbone's decoding throughput, so that neither modality monopolizes the shared per window budget.

Taken together, the two pathways supply the backbone with a compact, time-synchronized view of the surrounding scene at every window. These aligned visual and audio token streams are subsequently interleaved with the model's own outputs through the mechanism detailed in Section~\ref{jsp:2}, which is precisely what enables the front cerebellum to decide, chunk by chunk, whether to continue listening or to begin speaking.

\subsubsection{Streaming Chunk Flattening} \label{jsp:2}

To let the front cerebellum perceive and respond within a single autoregressive process, we adopt a flattened interaction formulation in the spirit of recent interaction models~\cite{thinkingmachines2026interactionmodels}, in which all modalities and both input and output are serialized into one causal token stream along a shared temporal axis. Under this view, the user request is no longer a privileged conversational role that gates generation. Instead, incoming speech and vision are treated as part of a continuously observed world state, and the model is situated in an always on environment where \textbf{it must decide not only what to produce but also whether and when to produce it in every chunk}.

Concretely, we partition the continuous interaction into fixed windows of one second and assemble one chunk per window. Each chunk is the flattened concatenation of three time-aligned segments: the encoded audio and visual tokens observed during that one-second window, a predicted control token, and the $N$ text tokens the model chooses to emit in response, where $N$ is determined by the model and may be zero. The interaction is then serialized by concatenating consecutive chunks into a single sequence that is consumed by a standard causal language backbone. Within each chunk, the model first attends to the newly arrived perceptual tokens and only then generates its output, so that every emitted token is conditioned on the most recent observation. Because the model reaches a decision once per window, proactive behavior arises naturally from the same mechanism and no external voice activity detection module is required to trigger responses. Marking the boundaries between the perceptual segment and the generated segment explicitly further helps the model distinguish observed inputs from its own outputs and stabilizes streaming decoding.

We manage context through a fixed budget of $128$ chunks, corresponding to a rolling temporal receptive field of roughly two minutes of interaction. As the dialogue proceeds, this window advances in a sliding manner: once the budget is reached, the oldest chunks are evicted as new ones are appended, so the effective context length stays bounded while the model retains the most recent and interactionally relevant history. This bounds context-dependent per step inference cost during long sessions and prevents unbounded growth of the flattened sequence.

The control token predicted at the head of each chunk's output segment governs the interaction and takes one of three values. 1) A \texttt{listen} token indicates that the model chooses to remain silent for the current window and continue observing, in which case the output segment carries no text and the chunk simply advances perception. 2) A \texttt{speak} token commits the model to generating spoken content in the current window, so that the subsequent $N$ text tokens are produced and passed downstream for speech synthesis. 3) An \texttt{interrupt} token allows the model to halt an ongoing utterance when the evolving context warrants it, for instance when the user begins speaking or the scene changes in a way that makes the in progress response stale. Predicting this discrete control decision before any content is generated decouples the question of whether to speak from the question of what to say, which we find yields more stable full-duplex behavior than entangling the two in a single prediction step.

\subsubsection{Speech Generation} \label{jsp:3}
Rather than having the LLM backbone emit acoustic units directly, the front cerebellum decouples semantic planning from acoustic realization and delegates waveform production to a pair of lightweight speech decoders. Forcing a language backbone to autoregress over speech tokens, which are typically emitted at a far higher frame rate than text, both inflates the number of decoding steps per second and tends to erode the model's core linguistic competence~\cite{xie2024miniomni2}. By keeping the backbone in the text domain and offloading acoustic modeling to compact downstream modules, the cerebellum sustains realtime spoken interaction at roughly the cadence of human speech while preserving the reasoning and instruction following behavior of the backbone. The generation pipeline proceeds in two stages: a speech token decoder that produces discrete speech units, followed by a streaming flow matching decoder that renders those units into an audio waveform.

\vspace{-0.5cm}

\paragraph{Speech token generation.} Natural speech requires not only correct pronunciation but also prosody, emphasis, and speaking style that are consistent with the surrounding context and the user's request. We take the final layer hidden state in the backbone, reshape it through a projection layer, and inject it with text token into a small autoregressive speech token decoder that emits the corresponding discrete speech tokens~\cite{du2024cosyvoice1}. Because prosodic and stylistic decisions are effectively pre-encoded in the backbone's hidden states, the speech decoder is relieved of high-level planning and can devote its limited capacity to fine-grained acoustic modeling. The discrete units themselves follow the supervised semantic token formulation of recent scalable speech synthesizers~\cite{du2024cosyvoice2}, which yields a single codebook, low bitrate representation that is well matched to autoregressive prediction. To keep the spoken output tightly coupled to the concurrently observed environment, the input text tokens and the output speech tokens are not generated in separate passes but interleaved along the shared timeline, following the time-aligned scheme described in Section~\ref{jsp:2}.

\vspace{-0.5cm}

\paragraph{Waveform synthesis.} The discrete speech tokens are converted into a continuous waveform by a streaming flow matching decoder~\cite{du2024cosyvoice2,cosyvoice3}. Conditioned on the reference audio carried in the multimodal system prompt, the decoder reconstructs the target mel-spectrogram from the semantic tokens through a conditional flow matching objective and then renders it to audio, which also endows the cerebellum with zero-shot voice control: the timbre and vocal identity of the synthesized speech are determined by the reference rather than fixed at training time. Critically, the decoder operates in a chunk-wise, causal manner so that waveform chunks are emitted incrementally as speech tokens arrive, rather than waiting for the full utterance to be decoded. This streaming design keeps the end-to-end latency of speech output low and allows the audio stream to be produced and played back continuously, which is a prerequisite for the full-duplex interaction the front cerebellum is designed to support.

\section{Data Construction Pipeline}\label{sec:data_construction}

Training Gander requires substantially different data from conventional turn-based multimodal models. \textit{Beyond general speech and multimodal understanding, the model must learn when to listen or speak, how to react to continuously evolving audio-visual context, how to coordinate with the back brain during long horizon agentic tasks, and how to maintain reliable interaction under complex real-world conditions.} To this end, we construct a large scale corpus comprising realtime speech interaction, audio-visual interaction, and agentic interaction data. We additionally incorporate robustness oriented and negative supervision data covering challenging acoustic and conversational conditions, multi-party scenarios, and cases where the model should remain silent or suppress unnecessary responses, improving its reliability under realistic interaction settings. The detailed composition and distribution of Gander's training data are summarized in Table~\ref{tab:gander_training_data}. The corpus is organized into four data families: speech interaction data, audio-visual interaction data, agentic interaction data, and robustness and negative data. We describe the composition and construction of each family in turn below.

\subsection{Speech Interaction Data}

Speech interaction data establish the \textit{basic capabilities} required for realtime spoken interaction and account for approximately 37\% of Gander's training corpus.
The corpus spans general dialogue, spoken instruction following and question answering, full-duplex interaction, and simultaneous speech translation.
Beyond semantic understanding and response generation, it supervises interaction behavior over continuously arriving speech, including turn-taking, interruption, overlapping speech, and response timing, thereby teaching the model when to listen, when to continue speaking, and when to yield the conversational floor as the interaction unfolds.
The full-duplex portion of this corpus denoted \textit{InteractionSpeech}, is produced by a dedicated pipeline that synthesizes interaction behavior explicitly~\cite{interactspeech}, rather than inheriting it from turn-based corpora in which interruption and overlap are absent by construction. This pipeline proceeds in four stages which described in turn below: we first collect dialogue content from two complementary \textit{sources}, then annotate the two \textit{interaction events} that define duplex behavior, \textit{render} the result to audio on an explicit timeline, and finally apply \textit{quality control} to the synthesized dialogues.

\textbf{Dialogue sources.} Content is drawn from two complementary sources. 1) The first is synthesis from topic seeds. We curate 11.2K scene and topic seeds spanning 45 everyday and task oriented scenarios, including education, healthcare, travel, financial and public services, and customer support, and expand each seed into a multi turn spoken dialogue with DeepSeek-V4-Pro~\cite{xu2026deepseek}. Generation is constrained to between 8 and 18 turns and at most 96 seconds of aggregate speaking time, which keeps every sequence within the streaming context of the front cerebellum and prevents the model from being trained predominantly on unnaturally long monologues. 2) The second source converts existing multi turn dialogues into duplex form, drawing on both real assistant interaction logs and public conversational corpora. Because such material is written to be read rather than spoken, every candidate first passes a spoken suitability filter that discards turns dominated by markdown structure, code, URLs, or mathematical notation, as well as turns whose length or information density would be implausible in speech. The filter rejects the majority of candidate dialogues; those retained preserve authentic user intent and phrasing, into which interaction events are subsequently injected. Each dialogue is further conditioned on one of five interaction profiles, namely constraint clarification, process control, user correction, failed service entry, and safety or urgency stop, ensuring that interruptions arise from diverse communicative motivations rather than from a single recurring pattern. English data are obtained both by direct generation and by translating validated Chinese dialogues. In the latter case only the marked dialogue text is translated, after which all derived fields are parsed again under English tokenization, so that interruption points are selected according to English word order rather than transferred from Chinese.

\begin{table*}[t]
    \centering
    \caption{
        Detailed distribution of the Gander's training corpus.
    }
    \label{tab:gander_training_data}
    \small
    \setlength{\tabcolsep}{4pt}
    \renewcommand{\arraystretch}{1.10}

    \begin{tabularx}{\textwidth}{
        @{}
        >{\raggedright\arraybackslash}m{2.25cm}
        >{\raggedright\arraybackslash}p{4.2cm}
        r
        r
        >{\raggedright\arraybackslash}X
        @{}
    }
        \toprule
        \textbf{Data family}
        & \textbf{Fine-grained category}
        & \textbf{Examples}
        & \textbf{Fraction}
        & \textbf{Main supervision} \\
        \midrule

        \multirow[c]{5}{=}{
            \textbf{Speech Interaction}
        }
        &
        Foundational dialogue
        &
        539.4K
        &
        20.00\%
        &
        General spoken dialogue.
        \\

        \cmidrule(lr){2-5}

        &
        Basic capabilities
        &
        26.9K
        &
        1.00\%
        &
        Basic conversational data covering math, instruction following, and logic.
        \\

        \cmidrule(lr){2-5}

        &
        InteractionSpeech
        &
        260.8K
        &
        9.67\%
        &
        Full-duplex turn-taking, interruption, and multi turn interaction.
        \\

        \cmidrule(lr){2-5}

        &
        Spoken question answering
        &
        165.1K
        &
        6.12\%
        &
        Short form spoken question answering.
        \\

        \cmidrule(lr){2-5}

        &
        Simultaneous speech translation
        &
        19.0K
        &
        0.70\%
        &
        Incremental bilingual speech translation.
        \\

        \midrule

        \textbf{Audio--Visual Interaction}
        &
        \begin{tabular}[t]{@{}l@{}}
            Streaming video QA \\
            Streaming video narration \\
            Proactive visual response
        \end{tabular}
        &
        1.1M
        &
        40.66\%
        &
        Streaming video QA, event grounding, narration,
        and realtime visual interaction.
        \\

        \midrule

        \multirow[c]{3}{=}{
            \textbf{Agentic Interaction}
        }
        &
        Audio agentic interaction
        &
        320.2K
        &
        11.87\%
        &
        Speech-driven task delegation and lifecycle interaction.
        \\

        \cmidrule(lr){2-5}

        &
        Multimodal agentic interaction
        &
        36.0K
        &
        1.33\%
        &
        GUI based multimodal agent interaction.
        \\

        \cmidrule(lr){2-5}

        &
        Tool assisted reasoning
        &
        3.4K
        &
        0.13\%
        &
        Tool-grounded reasoning over spoken STEM tasks.
        \\

        \midrule

        \multirow[c]{4}{=}{
            \textbf{Robustness and Negative Data}
        }
        &
        Irrelevant video robustness
        &
        116.2K
        &
        4.31\%
        &
        Robustness to irrelevant visual context.
        \\

        \cmidrule(lr){2-5}

        &
        No command environments
        &
        40.0K
        &
        1.48\%
        &
        Silence under non directed environmental input.
        \\

        \cmidrule(lr){2-5}

        &
        Anti interference
        &
        64.7K
        &
        2.40\%
        &
        Robustness to noise and overlapping distractors.
        \\

        \cmidrule(lr){2-5}

        &
        Multi party interaction
        &
        8.8K
        &
        0.33\%
        &
        Speaker and addressee tracking.
        \\

        \midrule

        \multicolumn{2}{l}{\textbf{Total}}
        &
        \textbf{2.7M}
        &
        \textbf{100.00\%}
        &
         \\

        \bottomrule
    \end{tabularx}
\end{table*}

\textbf{Interaction events.} Two events are annotated. In a \textit{competitive interruption}, the user barges in before the assistant finishes, and the assistant's remaining words form a hidden continuation that temporally overlaps the incoming user speech but is never heard by the user. In a \textit{supportive backchannel}, the user emits a brief acknowledgment while the assistant retains the floor and continues without pausing. The hidden continuation is retained as text but excluded from synthesis, so that the model observes exactly the acoustic evidence available to a real listener, and the subsequent user turn is verified not to reuse information it could not have heard. The distinction between the two events is enforced rather than assumed: a turn is accepted as a backchannel only if it is embedded within the interlocutor's ongoing utterance, falls below a short-length threshold of eight characters in Chinese or six words in English, and either matches the lexicon or is explicitly marked during generation. Candidates carrying interrogative, requestive, or negative-stance cues are demoted to ordinary turns, preventing genuine floor-taking utterances from being mislabeled as acknowledgments. Backchannel realizations are sampled from a bilingual lexicon of 266 Chinese and 170 English expressions, organized into 11 intent categories and conditioned on conversational context, which avoids the repetitive acknowledgments typical of small fixed word lists.

\textbf{Rendering and timing.} Only the user channel is rendered to audio, using voice cloning TTS~\cite{cosyvoice3,hu2026qwen3} over a pool of speakers, while assistant turns remain textual and occupy the timeline as duration placeholders. Every turn is assigned a global onset, a duration, and an overlap interval, so that the timing of interruptions and backchannels constitutes explicit supervision rather than an incidental artifact of concatenation. To avoid unstable synthesis, we emit no paralinguistic tokens and retain only two markers: the interruption marker, which delimits the overlap region, and the backchannel marker, which is stripped before synthesis.

\textbf{Quality control.} Filtering proceeds in two stages. A rule based gate first discards dialogues whose interruption points are linguistically implausible, whose overlapping segments are degenerate, whose interrupting turns leak content the user could not have heard, or whose interaction is left unresolved. Surviving samples are then rated by an LLM judge along four dimensions, namely naturalness, assistant coherence, interruption plausibility, and backchannel plausibility, and are retained only if they exceed a threshold on every applicable dimension. Dialogues obtained by rewriting undergo a further round of LLM based repair and curation targeting interaction and semantic quality. The resulting 260.8K dialogues distribute interruption onsets broadly across the interrupted turn rather than concentrating them near its end, and the majority of barge-ins begin directly with content rather than with a discourse marker, which prevents the model from associating interruption with a small set of stereotyped lexical cues.


\begin{figure}[t]
\centering
        \includegraphics[width=0.98\textwidth]{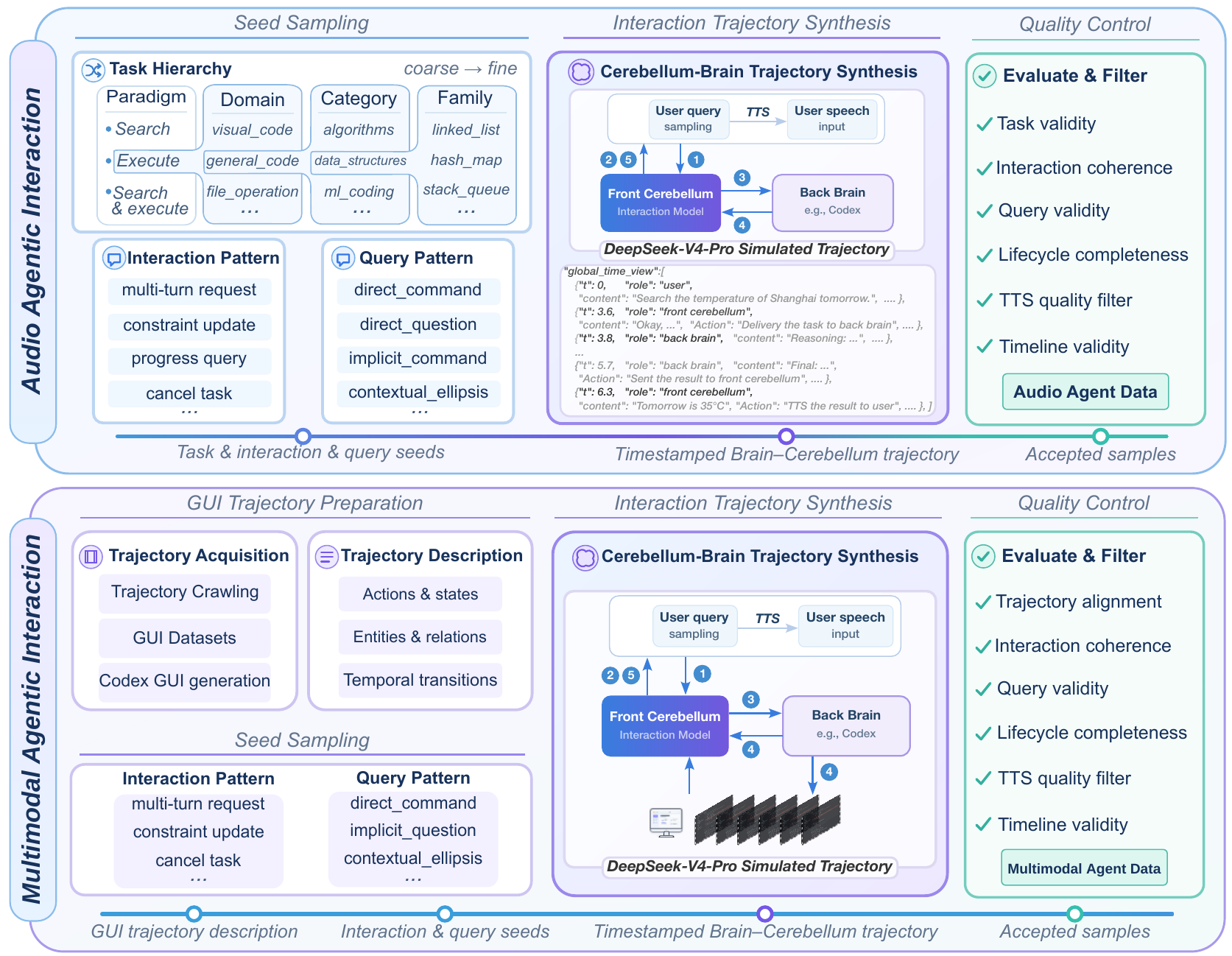}
\caption{Construction pipelines for agentic interaction data. Agent data are generated from hierarchically sampled task and interaction patterns, while multimodal-agent data originate from action-aligned video and GUI trajectories with structured descriptions. Both pipelines jointly synthesize user audio queries, front cerebellum responses, and back brain execution trajectories, followed by automatic evaluation and filtering.}
\label{fig:agentic_data_pipeline}
\end{figure}

\subsection{Audio-Visual Interaction Data}
Audio-visual interaction data extend realtime interaction from speech only conversations to continuously evolving multimodal environments, accounting for approximately 40\% of Gander's training corpus. 
The video interaction data are collected from JoyAI-VL~\cite{yao2026joyai}, LiveCC~\cite{chen2025live}, and Streamo~\cite{xia2026streaming}, and organized into three task categories: streaming video question answering, streaming video narration, and proactive visual response. All samples first undergo quality assessment and filtering, followed by temporal alignment refinement using Qwen3.5-297B-A17B~\cite{qwen3_5} to construct time-aligned targets and correct inaccurate alignments in the original data. This process yields approximately 1.1M high quality audio-visual interaction pairs. The target responses are further rewritten with DeepSeek-V4-Pro~\cite{xu2026deepseek} to normalize response lengths to an average rate of 8 tokens per second, balancing response informativeness and speech latency, while user utterances are synthesized into speech using Qwen3-TTS~\cite{hu2026qwen3}. The resulting data train the model to continuously integrate incoming visual evidence with the ongoing interaction, track evolving events and state changes, and ground its responses in information available up to the current moment, enabling timely responses when relevant visual events emerge.

\subsection{Agentic Interaction Data}
While the preceding data primarily focus on direct user interaction with the front cerebellum, agentic interaction data extend this setting to coordinated interaction among the\textit{ user, front cerebellum, and back brain}.
The agentic interaction corpus consists of three categories: audio agentic interaction, multimodal agentic interaction, and a small set of tool assisted reasoning data. The first two categories focus on long horizon agentic interaction, where the front cerebellum maintains continuous interaction with the user while coordinating with the back brain throughout asynchronous task execution. Tool assisted reasoning provides auxiliary supervision for tool grounded reasoning over spoken tasks. Unlike conventional tool use examples that primarily capture isolated tool invocation or task completion, these interaction trajectories interleave multi-turn user interaction with evolving backend states, execution progress, and intermediate results. They span the full task lifecycle, including task initiation, follow up requests, constraint updates, clarification, progress inquiries, cancellation, and final result delivery. The construction of audio agentic and multimodal agentic interaction data is described below, with their pipelines illustrated in Figure~\ref{fig:agentic_data_pipeline}. 

Audio-agentic interaction data are constructed through a seed-driven trajectory synthesis pipeline. Structured task seeds are first sampled at multiple levels, including task class, domain, category, and task family, together with diverse query and interaction patterns. The task space covers representative backend-agent workflows such as execution, information search, and search-then-execute, while the interaction patterns characterize how users may interact with an ongoing task. The back brain operates as an independent task executor, while the supervision focuses on how the front cerebellum communicates and coordinates around evolving task states. 
Given each seed, spoken user requests and corresponding interaction trajectories among the user, front cerebellum, and back brain are synthesized using DeepSeek-V4-Pro to capture the information exchange throughout task execution.
The resulting trajectories are evaluated and filtered for task consistency, interaction validity, and overall quality.

Multimodal-agentic interaction extends such coordination to visually grounded tasks, where the front cerebellum must continuously track evolving visual states while interacting with both the user and the back brain. The underlying GUI and video trajectories are assembled from three complementary sources: crawled interaction trajectories, existing GUI datasets~\cite{wang2026opencua}, and GUI trajectories generated with Codex, yielding approximately 36K examples in total. These trajectories contain temporally ordered observations and actions, which are converted by Qwen3.5-297B-A17B into structured descriptions of environment states, task progress, and interaction context. DeepSeek-V4-Pro is used to synthesize spoken user requests and corresponding interactions among the user, front cerebellum, and back brain, grounded in the observed visual states. These trajectories require the front cerebellum to follow evolving interface or scene states, communicate with the user, and remain synchronized with backend actions and intermediate results. Finally, the synthesized trajectories are evaluated and filtered for temporal consistency, task validity, and interaction quality.

\subsection{Robustness and Negative Data}
Reliable realtime interaction requires not only appropriate responses to valid user requests, but also robust response control under irrelevant, noisy, and multi-party conversational conditions.
Accordingly, robustness-oriented and negative supervision covers four settings: irrelevant visual context, no-command environments, acoustic and conversational interference, and multi-party interaction.
Irrelevant-video negative samples are constructed by pairing web-crawled videos with unrelated user instructions.
Irrelevant-video and no-command examples encourage the model to ignore unrelated events and remain silent when no valid user request is present.
Anti-interference data improve robustness to noise and overlapping distractors, while multi-party interaction data support speaker and addressee tracking to determine whether an utterance is directed toward the assistant.
Together, these data reduce spurious responses and improve interaction reliability in complex real-world environments.

\section{Experiments}
\label{sec:main_results}

We evaluate Gander along the three axes that its architecture is designed to serve:
full-duplex interaction with tool use, spoken conversational ability, and multimodal understanding
over audio-visual streams. The three differ in how much of the system they exercise, from
full-duplex interaction with tool use, which engages both tiers, to multimodal understanding,
which uses the front cerebellum answering in text. Read together they also serve a diagnostic
purpose, highlighting complementary strengths and helping relate remaining errors to
interaction training and the participating components. Because the settings differ in scoring protocol and in which components participate,
their scores are not commensurable and we report them separately.

\subsection{Evaluation Setup}
\label{sec:eval_setup}

\paragraph{Model under evaluation.}
All results are obtained from a single Gander checkpoint, with the Talker and the streaming
speech decoder attached wherever speech output is required. Decoding parameters and the streaming
unit format are identical across evaluations, and the front cerebellum's system prompt is
byte-identical to the one used during training, with no benchmark-specific tuning. Where the back
brain participates, it is instantiated training-free through the Codex worker provider of
Section~\ref{sec:e_bench}, driven by GPT-5.6, and may call only the tools the benchmark exposes.

\paragraph{Full-duplex interaction.}
We use Full-Duplex-Bench~v3~\cite{fullduplexbench_v3}, which places a spoken assistant in
tool-augmented service scenarios containing natural disfluencies, and evaluate all 100
scenarios. It reports tool-selection F1 (ToolSel), argument correctness (ArgAcc), whether the
spoken reply satisfies the user's intent (RespQual), a strict binary Pass@1 requiring the tool
multiset to be exactly correct with every argument right, and three interaction metrics: the
fractions of turns taken at an appropriate moment (Take-turn), begun before the user has
finished (Interrupt), and prefaced by a conversational placeholder (Filler). The last comes from
the suite's LLM-based latency analysis and is defined over the 91 of 100 scenarios that take the
turn without interrupting and for which that analysis returns a verdict. We run the released
scoring scripts unmodified, with the GPT-4o judge used by the baseline protocol, and evaluate
the deployed stack end to end: audio enters the Thinker, the Talker and speech decoder
synthesize a reply, and scoring reads an ASR transcript of that audio, so all reported numbers
include real speech-synthesis and recognition error. One asymmetry required
correction. The benchmark's system prompt instructs baselines to execute tools immediately
rather than ask clarifying questions, but the back brain, being an external agent, never
receives it; we therefore place a contract file carrying the same requirement in its working
directory so that both sides operate under equivalent instructions, leaving the front
cerebellum's prompt untouched. To separate the two tiers we additionally report a
back-brain-only condition that bypasses the front cerebellum and the audio path, driving the
same agent and tool server from the transcript of the user turn. Lacking an audio timeline, it
supports none of the three interaction metrics, so we leave those blank rather than report a
zero or a degenerate one.

\paragraph{Spoken conversation.}
We evaluate on the SpokenQA subsets Llama Questions and Web Questions and on the VoiceBench
subsets AlpacaEval and SD-QA~\cite{voicebench2025}, totalling 2{,}052 utterances, and adopt
the baseline table of Audio-Interaction~\cite{AudioInteractionModel} for comparison. Speech
enters the Thinker directly and the response it produces is scored as text; open-ended
responses are rated by a GPT-4o judge following the baseline protocol. Since the compared
systems differ in whether they must commit to their output while the user is still speaking, we
group them by interaction regime and read the comparison primarily within the full-duplex
group. Internal human evaluations additionally assess the naturalness and expressiveness of
spoken dialogue, complementing these automatic benchmark results.

\paragraph{Multimodal understanding.}
We evaluate on WorldSense~\cite{worldsense2025} and Daily-Omni~\cite{dailyomni2025}, comprising
4{,}369 multiple-choice questions. Prompts and the two-stage answer-extraction procedure
follow the upstream harnesses, answers are graded by exact letter matching with no LLM judge,
and questions whose answer cannot be extracted are counted as incorrect rather than discarded.
Decoding is greedy, so the evaluation is deterministic. This setting is turn-based: the whole
question is available before decoding begins, no speech is synthesized, and the back
brain is not involved, so it measures the front cerebellum's perceptual and reasoning ability
rather than its interaction behavior. To quantify the benefit of joint input
across modalities, we additionally run each question under three input conditions, presenting
audio and video jointly, video only, and audio only, for 13{,}107 inferences in total.

\subsection{Full-Duplex Interaction}
\label{sec:result_duplex}

\begin{table}[t]
    \centering
    \caption{
        Results on Full-Duplex-Bench~v3 (100 scenarios). The first four metrics are fractions,
        the last three percentages; baseline numbers are from the benchmark release. Bold marks
        the best value of each column, excluding the last row. $^{\dagger}$Text-driven and
        comparable to the cascaded pipeline rather than to the full-duplex rows; the three
        interaction metrics are not measurable without an audio timeline.
    }
    \label{tab:duplex_main}
    \small
    \setlength{\tabcolsep}{4pt}
    \renewcommand{\arraystretch}{1.10}
    \adjustbox{max width=\textwidth}{%
    \begin{tabular}{lccccccc}
        \toprule
        \textbf{Model}
        & \textbf{ToolSel}$\uparrow$
        & \textbf{ArgAcc}$\uparrow$
        & \textbf{RespQual}$\uparrow$
        & \textbf{Pass@1}$\uparrow$
        & \textbf{Take-turn}$\uparrow$
        & \textbf{Interrupt}$\downarrow$
        & \textbf{Filler}$\downarrow$ \\
        \midrule
        GPT-Realtime~\cite{openai_gpt_live_system_card_2026}
        & \textbf{0.876} & \textbf{0.680} & \textbf{0.792} & \textbf{0.600}
        & 96.0 & 13.5 & 16.9 \\
        Gemini Live 3.1~\cite{gemini_live_2026}
        & 0.817 & 0.588 & 0.718 & 0.540 & 78.0 & 19.2 & 31.7 \\
        Cascaded (Whisper\,$\rightarrow$\,GPT-4o\,$\rightarrow$\,TTS)~\cite{whisper}
        & 0.803 & 0.562 & 0.600 & 0.450 & \textbf{100.0} & 33.0 & 26.9 \\
        Grok~\cite{grok_voice_2026}
        & 0.797 & 0.542 & 0.617 & 0.430 & 94.0 & 25.5 & 44.3 \\
        Ultravox v0.7~\cite{ultravox2026}
        & 0.794 & 0.513 & 0.510 & 0.410 & 96.0 & 47.9 & 88.0 \\
        Gemini Live 2.5~\cite{gemini_live_2026}
        & 0.786 & 0.593 & 0.554 & 0.490 & 92.0 & 14.1 & \textbf{8.9} \\
        \midrule
        \textbf{Gander}
        & 0.759 & 0.503 & 0.490 & 0.400
        & \textbf{100.0} & \textbf{8.0} & 51.6 \\
        \midrule
        \textit{Gander, back brain only}$^{\dagger}$
        & \textit{0.934} & \textit{0.590} & \textit{0.740} & \textit{0.520}
        & --- & --- & --- \\
        \bottomrule
    \end{tabular}%
    }
\end{table}

Gander achieves the strongest turn-taking results in the table. It takes the floor at an appropriate
moment in all 100 scenarios, matched only by the cascaded pipeline, and begins speaking
prematurely in just 8.0\% of turns, against 13.5\% for GPT-Realtime and 47.9\% for the weakest
baseline. These results highlight a central strength of Gander's full-duplex design: responsive
turn-taking with a 9B interaction model across the evaluated service scenarios. On the four task-accuracy
metrics it trails, though by a narrow margin at the lower end of the table: Pass@1 is 0.400
against 0.410 for the weakest baseline and 0.600 for the strongest, with ToolSel and ArgAcc at
0.759 and 0.503 against that same baseline's 0.794 and 0.513.

The two timing metrics must be read jointly, because a system can trivially suppress
interruptions by waiting longer, at the cost of missing its turn altogether. Both failure modes
appear in the table: Gemini~Live~3.1 keeps interruptions to 19.2\% but answers in only 78.0\% of
scenarios, while the cascaded pipeline buys a perfect turn-take rate with a 33.0\% interruption
rate, illustrating the difficulty of identifying utterance boundaries when acoustic pauses do not
coincide with semantic completion. Gander combines both strengths, consistent with predicting
the interaction decision before any content is generated: the decision is conditioned on the same evolving representation that will produce the
response, so it can be deferred until the utterance is semantically complete rather than merely
acoustically quiet. Its 51.6\% filler rate is the one interaction metric on which it does not
lead. Filler marks the interval between taking the floor and delivering the answer, so unlike the
other two it does not describe when the model chooses to speak; during delegated tasks, brief acknowledgments can maintain
conversational continuity. The relatively high rate nevertheless leaves room to streamline
spoken progress updates and reduce unnecessary placeholders.

The accuracy gap is best read against the last row, which reports the back brain driven directly
from the user transcript. In that text-mediated regime, where the cascaded pipeline is the
appropriate comparison, it reaches Pass@1 0.520 against 0.450 and RespQual 0.740 against 0.600,
while its ToolSel of 0.934 exceeds every other system in the table, GPT-Realtime's 0.876
included. The execution tier thus retains strong tool-selection capability: the same agent and the same
tool set score above every baseline on tool selection once the front cerebellum and the speech
channel are removed from the path. Two effects separate that row from the end-to-end system.
The composite must decide for itself when to delegate, and it is scored on an ASR transcript of
synthesized speech rather than on text, which the RespQual drop from 0.740 to 0.490 is consistent
with. These runs do not separate the two effects, but point to delegation training and the speech
delivery path as promising targets for improving end-to-end task performance.

\subsection{Spoken Conversation}
\label{sec:result_voicechat}

\begin{table}[t]
    \centering
    \caption{
        Results on SpokenQA and VoiceBench, with systems grouped by interaction regime.
        Baseline numbers are taken from Audio-Interaction~\cite{AudioInteractionModel}.
        SpokenQA columns and SD-QA report accuracy (\%); AlpacaEval is rated on a 1--5 scale,
        and higher is better throughout. Within each group, \textbf{bold} marks the best and
        \underline{underline} the second-best value of each column.
    }
    \label{tab:voicechat_main}
    \small
    \setlength{\tabcolsep}{6pt}
    \renewcommand{\arraystretch}{1.10}
    \adjustbox{max width=\textwidth}{%
    \begin{tabular}{lccccc}
        \toprule
        \multirow{2}{*}{\textbf{Model}}
        & \multirow{2}{*}{\textbf{Size}}
        & \multicolumn{2}{c}{\textbf{SpokenQA}}
        & \multicolumn{2}{c}{\textbf{VoiceBench}} \\
        \cmidrule(lr){3-4} \cmidrule(lr){5-6}
        & & Llama Q. & Web Q. & AlpacaEval & SD-QA \\
        \midrule
        \multicolumn{6}{l}{\emph{Turn-based models}} \\
        Freeze-Omni~\cite{freezeomni2024}         & 7B & 72.00 & 44.73 & 4.14 & \underline{50.16} \\
        Baichuan-Omni-1.5~\cite{baichuanomni15}   & 7B & \textbf{78.50} & \underline{59.10} & \textbf{4.50} & 43.40 \\
        Qwen2-Audio~\cite{qwen2audio}             & 7B & 69.67 & 45.20 & 3.74 & 35.71 \\
        Qwen2.5-Omni~\cite{qwen25omni}            & 3B & 66.00 & 27.95 & 4.32 & 49.37 \\
        Qwen2.5-Omni~\cite{qwen25omni}            & 7B & \underline{75.33} & \textbf{62.80} & \underline{4.49} & \textbf{55.71} \\
        Phi-4-multimodal~\cite{phi4mm2025}        & 5.6B & 60.20 & 26.60 & 3.81 & 39.78 \\
        \midrule
        \multicolumn{6}{l}{\emph{Full-duplex streaming models}} \\
        Moshi~\cite{defossez2024moshi}            & 7B & 62.20 & 26.30 & 2.01 & 15.01 \\
        Audio-Interaction~\cite{AudioInteractionModel} & 3B & \underline{67.31} & \underline{54.34} & \textbf{4.28} & \textbf{52.14} \\
        \textbf{Gander}                           & 9B & \textbf{75.60} & \textbf{59.30} & \underline{3.96} & \underline{46.84} \\
        \bottomrule
    \end{tabular}%
    }
\end{table}

Gander leads the full-duplex group on both knowledge-oriented subsets of
Table~\ref{tab:voicechat_main}, reaching 75.60 and 59.30 on SpokenQA, ahead of Audio-Interaction
by 8.29 and 4.96 points and of Moshi by more than thirteen and thirty-three, and places second in
the group on the two VoiceBench subsets, 0.32 below Audio-Interaction on AlpacaEval and 5.30 below
it on SD-QA. We read the comparison within groups because the regimes are not equally
constrained: a turn-based model receives the utterance in full and may allocate arbitrary
computation before emitting anything, whereas a full-duplex model consumes it on a fixed
frame-level schedule and must additionally judge, at every step, whether the utterance is
finished. No metric here exercises that extra decision, which the preceding subsection measures
directly.

The result also holds against the turn-based systems. Across all nine systems, including
the turn-based baselines, Gander's SpokenQA scores still place second on
both subsets, 2.90 points behind Baichuan-Omni-1.5 on Llama Questions and 3.50 behind
Qwen2.5-Omni-7B on Web Questions, while AlpacaEval and SD-QA place sixth and fifth. That a
frame-synchronous model matches turn-based ones on spoken factual question answering is the main
point of the table: a streaming formulation can remain competitive on the knowledge-oriented tasks evaluated
here. The two weaker columns suggest complementary priorities for interaction training:
open-ended generation rated for overall response quality and a subset stressing accented
speech. Interaction supervision consists overwhelmingly of short conversational turns
emitted a few tokens at a time, which may provide less supervision for extended discursive composition and
broad acoustic coverage; the first is addressable by adding long-form
response supervision, the second by widening the accent distribution of the speech data.

The back brain remained available throughout this evaluation but was never invoked on any of the
2{,}052 samples, so every score in Table~\ref{tab:voicechat_main} is the front cerebellum acting
alone. That is the intended behavior: the routing policy escalates long-horizon tool work, which
the preceding subsection exercises, and not self-contained question answering. Such selectivity is
what makes a two-tier design more than an added latency cost, since indiscriminate escalation
would forfeit the compact tier's latency advantage. The corollary is that this table says nothing
about back-brain execution quality.

\subsection{Multimodal Understanding}
\label{sec:result_multimodal}

\begin{table}[t]
    \centering
    \caption{
        Multimodal understanding accuracy (\%) on WorldSense ($n$=3{,}172) and Daily-Omni
        ($n$=1{,}197). Baseline numbers are the officially reported
        ones~\cite{cui2026minicpm}; MiniCPM-o~4.5 is Gander's base model. Gander is evaluated
        in the audio-visual condition.
    }
    \label{tab:multimodal_main}
    \small
    \setlength{\tabcolsep}{6pt}
    \renewcommand{\arraystretch}{1.10}
    \begin{tabular}{lcc}
        \toprule
        \textbf{Model}
        & \textbf{WorldSense}
        & \textbf{Daily-Omni} \\
        \midrule
        Gemini 2.5 Flash   & 52.60 & 79.30 \\
        Qwen3-Omni         & 54.00 & 70.70 \\
        MiniCPM-o 4.5      & \textbf{55.70} & \textbf{80.20} \\
        \midrule
        \textbf{Gander}    & 49.62 & 78.53 \\
        \bottomrule
    \end{tabular}
\end{table}

\begin{table}[t]
    \centering
    \caption{
        Modality ablation on the same question sets, each condition presenting only the
        indicated input streams. Fusion gain is
        $\text{AV}-\max(\text{Video},\text{Audio})$.
    }
    \label{tab:multimodal_ablation}
    \small
    \setlength{\tabcolsep}{6pt}
    \renewcommand{\arraystretch}{1.10}
    \begin{tabular}{lcccc}
        \toprule
        \textbf{Benchmark} & \textbf{AV} & \textbf{Video} & \textbf{Audio}
        & \textbf{Fusion gain} \\
        \midrule
        WorldSense & 49.62 & 44.61 & 43.32 & $+5.01$ \\
        Daily-Omni & 78.53 & 59.40 & 57.81 & $+19.13$ \\
        \midrule
        Overall    & 57.54 & 48.66 & 47.29 & $+8.88$ \\
        \bottomrule
    \end{tabular}
\end{table}

Gander retains competitive multimodal understanding after interaction-oriented fine-tuning,
reaching 49.62 on WorldSense and 78.53 on Daily-Omni in Table~\ref{tab:multimodal_main}. On Daily-Omni
it stays within 1.67 points of its MiniCPM-o~4.5 initialization while remaining ahead of
Qwen3-Omni's 70.70. Since that initialization is itself the strongest of the three published
systems on both benchmarks, the comparison functions as a regression test against the model we
started from rather than a ranking against external baselines, and the retention is notable
alongside the model's new interaction capabilities. The interaction corpus includes streaming
video question answering, narration, proactive visual response, and negative examples, as
described in Section~\ref{sec:data_construction}. The vision tower is bit-for-bit unchanged, frozen throughout training
and absent from the resulting checkpoint, so the comparison reflects the base model's visual
representations combined with Gander's post-training multimodal processing and answering
behavior.

The asymmetry between the two benchmarks suggests that the effects of interaction training vary
across tasks. Daily-Omni falls by 1.67 points while WorldSense falls by 6.08, and the two differ
in what they demand, Daily-Omni testing temporal reasoning over audio-visual alignment and
WorldSense probing fine-grained perceptual attributes such as counting and localization. This
pattern is consistent with the emphasis on reasoning over jointly evolving streams in the
interaction corpus, and motivates balancing continuous interaction supervision with fine-grained
perceptual tasks. It identifies a concrete direction for extending Gander's multimodal strengths.

The ablation in Table~\ref{tab:multimodal_ablation} shows clear benefits from combining the two
streams rather than relying on a single modality. Presenting both together outperforms the better
single stream by 5.01 points on WorldSense and 19.13 on Daily-Omni, and neither benchmark is
carried by one modality: video and audio alone land within 1.3 points of each other on WorldSense
and within 1.6 on Daily-Omni, so the joint condition is not simply tracking whichever stream is
stronger. The size of the gain follows what each benchmark demands. Daily-Omni tests temporal
reasoning over audio-visual alignment, where neither stream is sufficient alone, and gains 19.13
points; WorldSense probes fine-grained perceptual attributes that are often available from a single
modality, and gains 5.01. These complementary gains highlight the value of retaining both streams within the
interaction model.

\section{Conclusion and Limitation}

In this work, we present Gander, a \textbf{\textit{duplex interaction model}} with an \textbf{\textit{asynchronous agent loop}}, built upon a Brain–Cerebellum framework and a streaming chunk-based paradigm. Gander brings together multimodal understanding, realtime interaction, and long horizon agentic execution. Several limitations and directions for future research remain.

\begin{itemize}

\item \textbf{Data and Model Scaling.} Our experiments suggest that agent invocation and conversational behavior remain sensitive to training data distribution, especially in complex multimodal scenarios. \textit{Scaling data and model capacity is a promising direction for improving the robustness and generalization of interactive and agentic capabilities}.
\item \textbf{Stable Post Training.} The current Gander model has not yet been extensively optimized with on-policy distillation~\cite{xopd,x3opd} (OPD) or reinforcement learning~\cite{shao2024deepseekmath,chen2026wavalign,ji2025wavreward,dualreasoner} (RL) for long horizon duplex interaction agent scenarios. Effective reward design, credit assignment, and optimization stability remain open challenges, particularly under joint multimodal interaction and agentic execution. The Brain–Cerebellum architecture further introduces new considerations for coordinating local interaction quality with global task objectives, motivating the development of more stable and scalable post-training methods.

\item \textbf{Brain–Cerebellum Architectural Exploration.} The Brain–Cerebellum framework enables low latency streaming interaction and long horizon reasoning. The current design primarily relies on ASR derived signals for communication between the two components. Future work should explore richer and more structured bidirectional communication, including improved information transfer from the Brain to the Cerebellum and more effective feedback from the Cerebellum to the Brain, to better coordinate realtime interaction and long horizon reasoning.

\item \textbf{Memory and Long Context Management.} Multimodal agentic interaction involves long multimodal histories, tool calls, and evolving task states, posing challenges for long horizon context and memory management~\cite{xie2026voicemem}. Developing efficient mechanisms for retaining and retrieving task-relevant information over extended interactions is an important direction for future work.

\item \textbf{Evaluation.} Existing benchmarks largely evaluate multimodal understanding, duplex interaction, and agentic execution in isolation, leaving the unified duplex interaction agent setting insufficiently evaluated. In addition, existing benchmarks do not adequately capture Brain–Cerebellum collaboration, including inter component communication and coordination. Further evaluation of sustained user--agent collaboration would complement the current benchmarks and human assessment of conversational naturalness.

\end{itemize}

Overall, Gander provides a foundation for a unified duplex interaction agent architecture. Further progress toward reliable long horizon deployment depends on advances in scaling, post training, Brain–Cerebellum coordination, memory, and evaluation, especially for sustained realtime interaction and complex task execution.

\bibliography{ref}



\end{document}